\documentclass[a4paper,12pt]{extarticle}
\usepackage{mathptmx}
\usepackage{setspace}
\usepackage[utf8]{inputenc}
\usepackage{xcolor}

\usepackage[margin=1in]{geometry}

\usepackage{graphicx}
\usepackage{standalone}
\usepackage{amsfonts}
\usepackage{amsmath}
\usepackage{amssymb}
\usepackage{tikz-cd}
\usepackage{blindtext}
\usepackage{enumitem}
\usepackage{braket}
\usepackage{mathtools}
\usepackage{environ}

\usepackage{stmaryrd}
\usepackage{tikz-cd}
\usepackage{xcolor}
\usepackage{pgfplots}
\usepackage{multirow}
\usepackage{tikz}
\usepackage{colortbl}
\usepackage{comment}
\usepackage{verbatim}
\usepackage{caption}
\usepackage{booktabs}
\usepackage{float}
\usepackage{subcaption}
\usepackage{fdsymbol}
\usepackage{bbm}
\usepackage{lscape}
\usepackage[blocks]{authblk}
\restylefloat{table}

\DeclareFontFamily{OT1}{pzc}{}
\DeclareFontShape{OT1}{pzc}{m}{it}{<-> s * [1.10] pzcmi7t}{}
\DeclareMathAlphabet{\mathpzc}{OT1}{pzc}{m}{it}

\allowdisplaybreaks

\newtheorem{theorem}{Theorem}[section]

\newtheorem{lemma}[theorem]{Lemma}

\newtheorem{definition}[theorem]{Definition}
\newtheorem{example}[theorem]{Example}
\newtheorem{remark}[theorem]{Remark}
\newtheorem{condition}[theorem]{Condition}
\let\olddefinition\definition
\let\oldexample\example
\let\oldremark\remark
\let\oldcondition\condition

\renewcommand{\definition}{\olddefinition\normalfont}
\renewcommand{\example}{\oldexample\normalfont}
\renewcommand{\remark}{\oldremark\normalfont}
\renewcommand{\condition}{\oldcondition\normalfont}
\newenvironment{proof}{\noindent{\bf Proof:}}{$\hfill \Box$ \vspace{10pt}} 

\newcommand\numberthis{\addtocounter{equation}{1}\tag{\theequation}}
\graphicspath{{Figures/}}
\providecommand{\keywords}[1]
{
  \small	
  \textbf{\textit{Keywords:}} #1
}

\usepackage{natbib}
 \bibpunct[, ]{(}{)}{,}{a}{}{,}%

\PassOptionsToPackage{hyphens}{url}
\usepackage[colorlinks,citecolor=blue,urlcolor=blue,hypertexnames=true]{hyperref}

\title{Influencing Product Competition Through Shelf Design}

\author{Francisco Cisternas$^{1}$, Wee Chaimanowong$^{2}$, Alan L. Montgomery$^{3}$,\\ Timothy Derdenger$^{4}$\\
    \small $^{1}$Assistant Professor at CUHK Business School, The Chinese University of Hong Kong, 12 Chak Cheung Street
Shatin, N.T., Hong Kong., +852 3943-3231, (fcisternas@cuhk.edu.hk) \\
    \small $^{2}$ PhD Student, CUHK Business School, The Chinese University of Hong Kong, 12 Chak Cheung Street, +852 6274-9422, (chaimanowongw@gmail.com) \\
    \small $^{3}$Professor of Marketing at the Tepper School of Business, Carnegie Mellon University, 5000 Forbes Avenue
Pittsburgh, PA 15213, 412 268-4562 (alanmontgomery@cmu.edu)\\
    \small $^{4}$Associate Professor of Marketing and Strategy at the Tepper School of Business, Carnegie Mellon University, 5000 Forbes Avenue
Pittsburgh, PA 15213, 412 268-9812 (derdenge@andrew.cmu.edu)}
\date{ \footnote{We would like to acknowledge that our data came from the Micro-Marketing Project at the University of Chicago, which was directed by Stephen Hoch.  Many researchers participated in the design and analysis of the shelving experiments used in this paper including Xavier Dr\`eze, Byung-Do Kim, Mary E. Purk, and Peter E. Rossi.  This research would not have been possible without their efforts and we would like to acknowledge their effort. Francisco Cisternas is the corresponding author.} \\ \small \today }

\pgfplotsset{compat=1.16}

\begin{document}
\pagenumbering{gobble}
\maketitle
\newpage
\renewcommand{\thefootnote}{\fnsymbol{footnote}}
\begin{titlepage}
\centering
\vskip 60pt
\LARGE Influencing Product Competition Through Shelf Design \par
\vskip 2em
\begin{abstract}

Shelf design decisions strongly influence product demand. In particular, placing products in desirable locations increases demand. This primary effect on shelf position is clear, but there is a secondary effect based on the relative positioning of nearby products. Intuitively, products located next to each other are more likely to be compared having positive and negative effects. On the one hand, locations closer to relatively strong products will be undesirable, as these strong products will draw demand from others---an effect that is stronger for those in close proximity. On the other hand, because strong products tend to attract more traffic, locations closer to them elicit high consumer attention by increased visibility. Modifying the GEV class of models to allow demand to be moderated by competitors' proximity, these two effects emerge naturally. We found that although the competition effect is usually stronger, it is not always the dominating effect.
Shelf displays can achieve higher profits by exploiting the relative influence on competition from shelf design to shift demand to higher profitability products. In the paper towel category, we found profitability differences of up to 7\% and displays with 3\% higher gross profits over the best shelf design present in our data.

\end{abstract}
\keywords{Shelf Design, Competition, Choice model, Spatially Correlated Logit Model}
\end{titlepage}
\renewcommand*{\thefootnote}{\arabic{footnote}}

\pagenumbering{arabic}

\section*{Introduction}


Retail shelf display is a defining point of contact between the consumer and retailer, since it is at this point that the purchase decision is made. During the purchase occasion, consumers face two main tasks: information acquisition and information integration. Shelf designs affect these tasks through visual cues such as product display, assortment, shelf attractiveness, or space allocation. Notice that shelf design could be considered either an attribute of the product or a factor of the environment that influences consideration. Shelf design can simplify the task of finding a routinely purchased item by placing the item front and center. Alternatively, it can make shopping more difficult by disbursing popular items throughout the shelf.

Our hypothesis is that there is a positive (or negative) effect of being close to competitors if the product has a relatively larger (or smaller) value to consumers. An important element in this argument about the value of shelf display is its distinction between absolute and relative positional effects. Absolute effects correspond to shelf positions that provide a gain regardless of the relative merits of the product. We conjecture that these absolute positional effects are due to higher visibility or increased traffic locations and are independent of the attributes of the product. But we posit that relative positional effects derive from being near competing brands. These competitive effects depend upon the relative attractiveness of other products in the spatial neighborhood, and could be positive because they draw upon the attractiveness of the competing products---or perhaps negative because they put a product in a more competitive environment. The consequence of these visibility and competitive effects is that shelf design can moderate product competition through spatial effects.

This research proposes a \textit{new} choice model that captures these price competition effects induced through shelf position, which has not received attention in the marketing literature. Our proposed choice model has better fit in all our experimental data, and better performance predicting market shares, when compared to traditional benchmarks. Moreover, we find that ignoring this competition effect causes a systematic bias in price sensitivity, overestimating its effect, and thus affects pricing and promotions strategies. Intuitively, ignoring the relative position of products forces the model to incorporate part of this omitted effect in other parameters, such as price sensitivity.

Our model permits cross price-elasticities to vary with respect to location and neighborhood, unlike traditional models. An analysis of these changes identifies display choices that increase overall shelf profits. From this display analysis, we generate recommendations for retail managers to improve the profitability of displays. Applying these recommendations to the paper towel category case yields a profit increase of roughly 5$\%$ relative to the existing shelf design.  

The contribution of this research can be summarized as follows. First, we make a methodological contribution whereby we develop a model that captures the effect of product display on product competition. Analysis of the elasticities from the implied choice probabilities show potential bias in price sensitivity when shelf-induced competition is ignored. Second, we deploy our model empirically with experimental data from the paper towel market sold in Dominick's Finer Foods stores in the Chicago area \textit{and} from a virtual store. We show that our model outperforms traditional benchmarks in terms of fit and predictive power, and confirm a price sensitivity bias. Third, using counterfactual evaluation of alternative shelf designs we find profit improving designs. Finally, we provide general recommendations for store managers based on our findings.

\section*{Literature Review}

Early research in shelf design focused on reducing operational costs (\cite{pauli1952better}, \cite{cox1970effect}, and \cite{anderson1979analysis}). \cite{bultez1988sh} developed a decision tool (S.H.A.R.P.) to help managers optimize shelf space allocation that was subsequently widely used in the industry. \cite{dreze1994shelf} assumed that there are fixed shelf position effects. \cite{corstjens1981model} and \cite{corstjens1983dynamic} create an adaptive spatial structure, but do not assume interactions between product attributes and shelf locations. More recently, \cite{hwang2005model}, \cite{rabbani2018profit}, \cite{bianchi2018allocating} and \cite{smirnov2019shelf} have included demand elasticities within the space allocation optimization problem. Although these demand models involve both neighborhood (cross space effect) and spatial (location effect) components, the cross space elasticity is assumed to be independent of relative position. Our conclusion is that researchers have considered spatial effects but have not considered how these spatial effects interact with the attractiveness of the products, which we believe to be an important aspect of this problem.

When consumers choose a product from a shelf display they gather information from the display and engage in comparisons before making a purchase decision. Eye-tracking studies (\cite{scekic2018product}, \cite{chandon2009does}, \cite{wedel2008review}) all point to the influence of arrangement or shelf design on choice. \cite{chen2021understanding} found that products being displayed at eye-level or end-caps result in higher sales. Similarly, \cite{bucklin2003model} reported that position influences clicks on websites, and \cite{agarwal2011location} found that advertising click-through rates decrease with position. 

This research points to location influencing demand as a consequence of consumer search behavior. Furthermore, products located near each other are more likely to be compared. \cite{mcgranaghan2019lead} found a spillover effect of promotions onto subsequent offers when the value of the lead offer was high. \cite{sayman2002positioning} argue that private labels benefit from being positioned\footnote{Their use of position is in a perceptual space and not necessarily shelf location.} near national brands. Their argument is that positioning store labels near national brands allows these weaker store brands to appeal to customers of the national brands using comparative advantages like lower price. The national brand draws consumers to it in a way that the store brand does not. More generally, we conjecture that there is a potential positional effect of being close to competitors. This effect depends upon the relative advantage of the product to the proximity of the competing products, where we measure proximity through shelf distance. We argue that this spatial effect is not just between national brands and private labels, but affects all products. An alternative justification for spatial competition comes from the context effect literature (\cite{rooderkerk2011}). Potentially, neighboring products on the shelf may alter the attractiveness of the focal product, which yields relative positional effects. 

\section*{Shelf Experiment Data} \label{sec:data}

Our dataset comes from the Micro-Marketing Project at the University of Chicago conducted between 1989 and 1994. The project was sponsored by numerous CPG manufacturers, and Dominick’s Finer Foods (DFF) which provided data and made their stores available to conduct controlled, field experiments. There are 25 categories--but not all participated in all phases of the project (there were less than 20 categories in the shelf experiments). There were three phases of the project: the first focused on shelf experiments, the second on category level price and promotional experiments, and the third focused on coordinating category level prices to understand store level effects. This has been a popular dataset for studying price and promotional effects, both because it is in the public domain and the carefully controlled price and promotion experiments. The reasons that we revisit this dataset is because it has an extensive set of controlled shelf experiments that have largely been ignored. Only one published study (\cite{dreze1994shelf}) used the shelf-experiments data, but their analysis was limited to estimating space elasticities and not understanding how price interacts with shelf design. Furthermore, these are high quality field-experiments since the stores were randomly assigned treatments, the experiments ran for many months, and they were checked in the field by research assistants to make sure the designs were followed. The experiments are older but we would point out that supermarket layout, shelf design, and the products analyzed (like paper towels) have not been altered during this time.  Hence, we are able to leverage this unique dataset to more rigorously study shelf design and contribute a new methodological approach to the analysis of shelf design.

We focus on the paper towel market to illustrate our approach and off several reasons for this choice. The paper towel category is a prototypical one with distinct quality tiers: premium national brands, a second tier of national brands, and a lower quality tier of private labels. The category is mature and frequently purchased (approximately once every two weeks). There are a relatively small number of products, and brands have few product versions: multi-roll or single-roll. The products are physically large, and the entire shelf display may occupy a whole aisle. This large size requires shoppers to walk the length of the aisle to make a selection, which makes shelf design effects more pronounced.. 

\textbf{Data Description.} Our dataset comprises 60 stores covering the greater Chicago area in which 26 products are displayed in every store over 16 weeks from July 11th through October 30th, 1991. The category is composed of three quality tiers: The premium national brands are Bounty and Brawny, the private label is Dominick's store brand, and the remaining products are associated with other national brands. We provide summary statistics of these products over the course of the data period below in the appendix (Table \ref{tab:ptw_desc}).  

In this category, a premium national brand product is not the top market share holder. Rather, Dominick's private label product holds an 11\% market share over other products from national brands such as Hi Dri, with its print paper towels, at 8.2\% and Mardi Gras Towels, at roughly 7\%. However, if we aggregate by brand, national premium brands do rank number one and two, respectively. Bounty holds 19.58\% followed by Scott at 17.52\%, HiDri at 11.86\%, Dominick's at 11.14\%, and Viva at 11.13\%. This market is competitive, with an HHI score\footnote{Herfindahl–Hirschman index Measures the industry concentration. An HHI below 1,500 indicates an unconcentrated industry} of 1,237. Additionally, we determine a strong negative correlation (-0.42) between price and market share as well as profit margin and market share (-0.46). 

\textbf{Experimental Shelf Designs.} Unfortunately, the summary statistics in Table \ref{tab:ptw_desc} do not provide any insight into the impact of shelf position or shelf competition on sales and market shares. In order to glean insight into these effects, we leverage an in-store experiment. The shelf experiment had three main objectives. The first was to determine if it was possible to move customers up in quality by organizing brands by quality and size versus only by brand to make it easier to "trade-up." The second objective was to hide unflattering price comparisons by increasing the difficulty of comparing regular, jumbo, and multi-packs.  The final purpose was to determine which shelf was viewed as high visibility in the paper towel category, by placing premium single rolls in highly visible locations to increase margins over the entire category.

The experimental design was straightforward, comprising three experimental shelf designs and a control design. The first treatment group organized products by quality and size by placing premium single rolls on the top shelf (providing high visibility), with multi-counts in the middle and low price brands on the bottom shelf. The second treatment organized shelves by placing low and mid price single rolls on the top shelf with multi-packs in the middle and premium on the bottom. The last treatment group made price comparisons very difficulty by vertically merchandising single rolls and multi-packs. The base shelf configuration merchandised all sizes within brand blocks. We illustrate each of these treatment and control groups with planograms below in Figure \ref{fig:planograms}.  

\begin{figure}[!ht]
    \begin{subfigure}[t]{0.4\textwidth}
    \centering
    \includegraphics[scale=0.25]{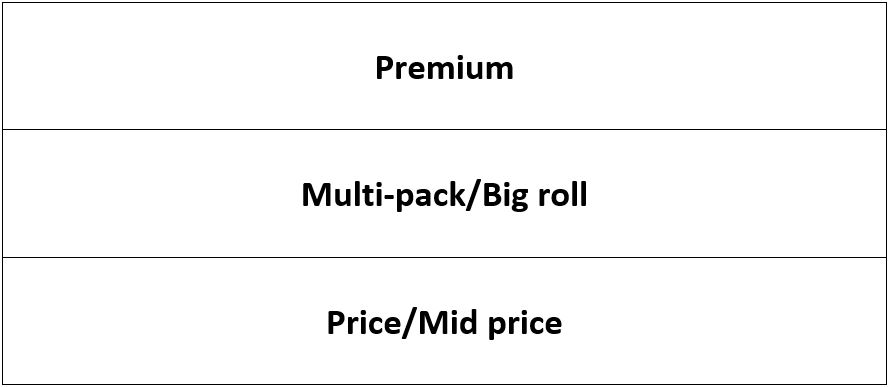}
    \caption{\footnotesize{Shelf treatment 1: Quality Tier 1. Products organized by quality and size with single rolls on the top shelf, multi-rolls in the middle, and lower price brands on the bottom.}}
    \label{fig:ptw_design1}
    \end{subfigure}\qquad\qquad\begin{subfigure}[t]{0.4\textwidth}
    \centering
    \includegraphics[scale=0.25]{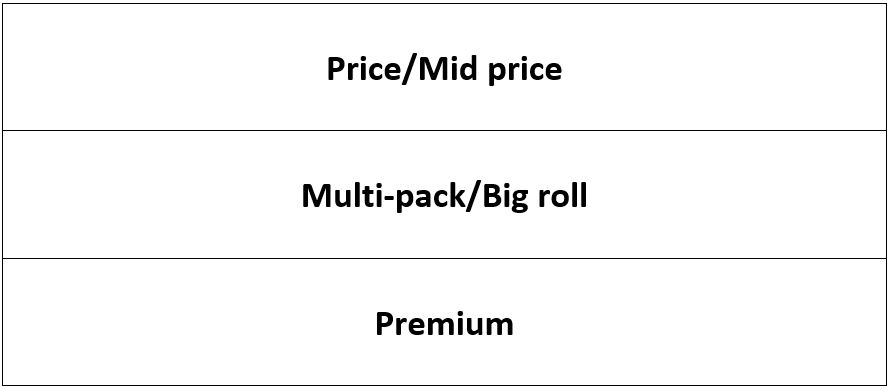}
    \caption{\footnotesize{Shelf treatment 2: Quality Tier 2. Organizing by quality and size, placing price/mid-price and single rolls on the top shelf, multi-rolls in the middle, and premium on the bottom}}
    \label{fig:ptw_design2}
    \end{subfigure}\qquad\qquad\begin{subfigure}[t]{0.4\textwidth}
    \centering
    \includegraphics[scale=0.25]{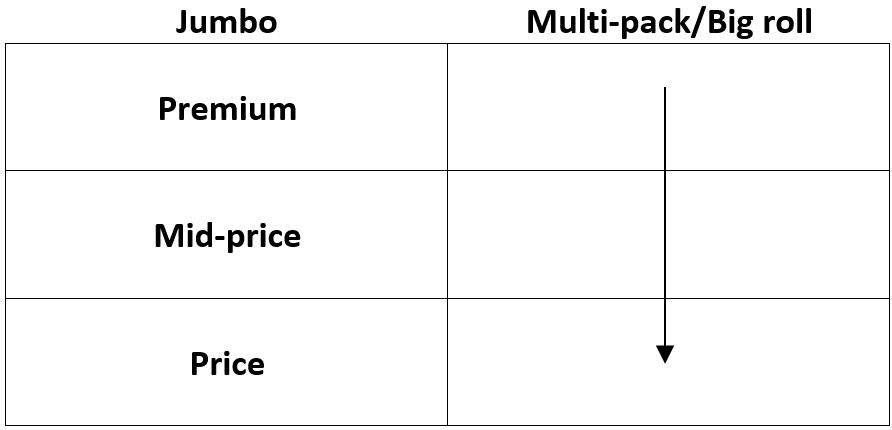}
    \caption{\footnotesize{Shelf treatment 3: Increased difficulty in shopping. Single jumbo size towels separated from the multi-rolls, the jumbo size organized by quality from top to bottom, and the multi-rolls organized by size from top to bottom}}
    \label{fig:ptw_design3}
    \centering
     \end{subfigure}\qquad\qquad\begin{subfigure}[t]{0.4\textwidth}
    \centering
    \includegraphics[scale=0.25]{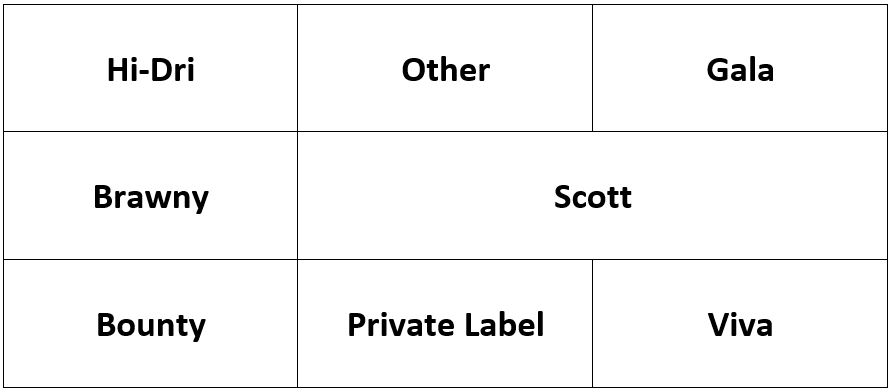}
    \caption{\footnotesize{Control shelf design. Products grouped by brand and premium brands are scattered through the design.}}
    \label{fig:ptw_control}
    \end{subfigure}
    \caption{\small{Shelf Design Planograms}}
    \label{fig:planograms}
\end{figure}

To execute the experiments, 60 stores were randomly assigned to one of these three treatments or the control. The designs and planograms were developed in conjunction with Procter and Gamble. Data was collected from July 11-October 30 1991. Additionally, stores were visited weekly to verify that shelf designs were implemented and to flag any potential problems (e.g., missing products or departure from shelf design).

\subsection*{Exploratory Data Analysis} \label{sec:model_lit}


In our analysis, variation in shelf position is key to determining whether there are absolute shelf position effects and relative competitive effects due to shelf position. Unfortunately for many datasets, shelf design is fixed; therefore, it is nearly impossible to separate these two effects from each other because there is no variation in the data with respect to shelf position. However, the shelf design experiments make it possible though the variation in shelf design that were introduced. Before developing a formal model, we explore below the results of the field experiment on shelf design from Dominick's Finer Foods.



\textbf{Data Suggesting Positional Effects.} 
 We present a simple analysis by way of ANOVA to determine whether shelf position impacts sales and how neighboring products impact those sales. 
 First, an ANOVA test indicates significant\footnote{The F values associated with horizontal and vertical locations are 160.3 and 51.75, respectively, and the p-value associated is less than 2e-16 in both cases.}  differences in horizontal and vertical product location influence on SKU-level movement after controlling for individual store, week, and product intercepts, which signal evidence for a more established absolute shelf effect. Clearly the shelf position matters. Items on the middle shelf do best, followed by the top shelf, and then the lower shelf. This can be inferred when we evaluate the model using higher order terms of the horizontal location, where the linear term is negative, the second order is positive, and the third order is negative. For the vertical location, on the other hand, the coefficient associated to the mid level is positive and significant.
 
\begin{figure}[ht!]
     \centering
    \begin{subfigure}[b]{0.5\textwidth}
         \centering
         \includegraphics[width=\textwidth]{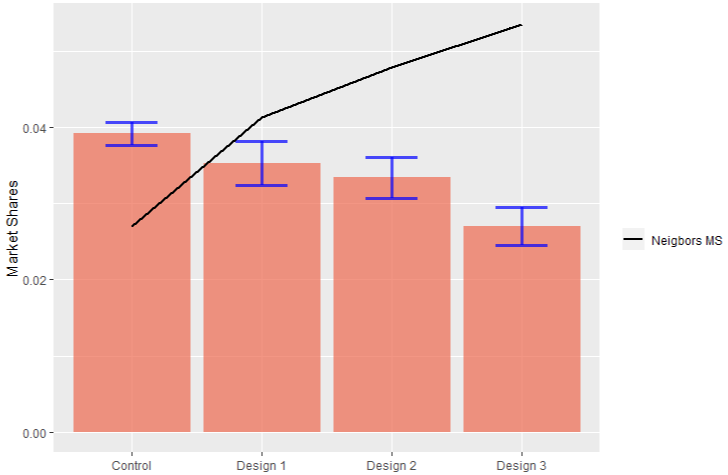}
         \caption{\footnotesize{Hi Dri's Market Shares in all 4 designs.}}
         \label{fig:model_free_a}
    \end{subfigure}
    \hspace{0.1cm}
    \begin{subfigure}[b]{0.22\textwidth}
         \centering
         \includegraphics[width=\textwidth]{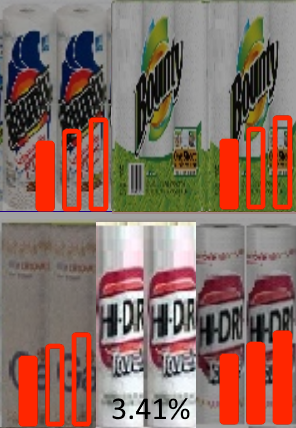}
         \caption{\footnotesize{Design 1}}
         \label{fig:model_free_b}
    \end{subfigure}
    \hspace{0.5cm}
    \begin{subfigure}[b]{0.2\textwidth}
         \centering
         \includegraphics[width=\textwidth]{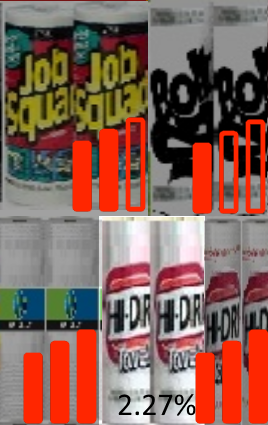}
         \caption{\footnotesize{Design 3}}
         \label{fig:model_free_c}
    \end{subfigure}
    \caption{\footnotesize{Hi Dri's market shares across shelf designs. In (a) neighborhood strength is shown as average market shares of immediate neighbors as a solid black line. On (b) and (c) specific neighborhoods of design 1 and 3 are shown and the corresponding strength is show with temperature bars.}}
    \label{fig:model_free}
\end{figure}

 Regarding the relative effect, we find that products tend to have smaller market shares when immediate neighbors are popular products with high average market shares. Indications of these effects can be seen on both products from high and low quality tiers, but stronger in unpopular products. Figure \ref{fig:model_free_a} shows the demand of the unpopular product Hi Dri. Demand is highest in the control design, where the average market share of immediate neighbours is low (black line). As the average market share of competition increases in the experimental designs, the demand for Hi Dri decreases. In Figures \ref{fig:model_free_b} Hi Dri has a market share of 3.41\% and is surrounded by lower quality products, except for a high quality product to its right (indicated by the temperature bars overlaid on the product image). Yet, in Design 3 (Figure \ref{fig:model_free_c}), where the shelf position is the same for Hi Dri but the surrounding competing products' quality increases, the market share of Hi Dri decreases to 2.27\%.
 
 In summary, this exploratory analysis suggests that both absolute and relative effects of the shelf position can significantly affect the demand. 

\section*{The NCL Model}  \label{sec:model_defn}
The next challenge is to specify a model of product choice that encompasses both our notion of absolute and relative shelf effects. Most choice models are derived from the random utility maximization (RUM) hypothesis. This is the approach that we propose as well. The multinomial logit model (MNL) is by far the most popular choice model. It follows as a consequence of the distributional assumptions on the random component being type I extreme value that is independently and identically distributed (\textit{iid}). The key advantage of the MNL model is its elegant and closed form solution for expressing choice probabilities, which facilitates estimation and predictions that support managerial decisions. Simplistically we could introduce fixed effects in a MNL model for shelf design that measure attractiveness of shelf locations, but such a model could not capture relative shelf design effects.

A well know limitation of the MNL is the independence of irrelevant alternatives (IIA). For a discussion see \cite{luce1965utility} and \cite{1973conditional}. In our context IIA is problematic since as products are added within a shelf location products they may draw share differentially from those that are close but not based upon overall attraction. Many extensions to the MNL model have been proposed to circumvent IIA, but the one we follow is a relaxation on the error term to allow more flexible error structures. One approach is the use of a nested logit (NL) derived by \cite{williams1977formation}, \cite{daly1976zachary}, \cite{ben1977disaggregate}, and \cite{mcfadden1981econometric}. 

In our problem a nested structure could capture competition within an limited shelf area of partition, which would address the IIA problem. However, the nested logit structure does not permit general spatial competition that would better capture our notion of relative competition from neighboring products. The generalized extreme value model (GEV) derived by \cite{1978modeling} and generalized by \cite{salomon1983use} introduces correlation into the error terms that relaxes the \textit{iid} assumption. \cite{small1987discrete} introduced the ordered GEV (OGEV) to account for ordered-choices like number of cars to own. Later extensions allow alternatives to belong “fractionally” to multiple nests, such as the paired combinatorial logit (PCL) by \cite{chu1989paired} and extended by \cite{koppelman2000paired}, cross nested logit (CNL) by \cite{vovsha1997application}, product differentiation logit (PDL) by \cite{bresnahan1996market}, and the multinomial logit order GEV (MNL-OGEV) by \cite{bhat1998analysis}.

Further flexibility in the correlation structure was suggested by \cite{wen2001generalized} with their generalized nested logit (GNL) formulation as well as a similar formulation by \cite{swait2001choice} with a latent choice set generation logit (GenL). The added flexibility of these models allowed general patterns of cross-substitution to be captured, but at the cost of greater complexity. Along these lines \cite{bhat2004mixed} introduced the spatially correlated logit model (SCL) for transportation choices in which adjacent geographical neighborhoods areas were considered correlated. Later \cite{sener2011accommodating} extended this idea proposing the generalized spatially correlated logit (GSCL) to account for unobserved spatial correlation. 


Our research follows these formulations to accommodate both absolute and relative effects of shelf position. Specifically, the correlation structure is a function of the shelf design such that closer neighbors having greater correlation. We refer to this model as the \textit{Neighborhood-induced correlation logit model} (NCL). While the roots of our model are in the choice literature, it also is related to the spatial literature (\cite{cressie2015}, \cite{bradlow2005}). Many spatial models tend to focus on macro-spatial effects that correspond with geographic entities like zip codes (\cite{bellsong2007}). These spatial effects are similar to our approach. However, the interpretation of these neighborhoods are quite distinct from ours. This is not to say research on context effects that consider spatial characteristics has not been considered. Specifically, \cite{dotson2018} consider a probit model with a covariance structure that depends upon the similarity of brands in a preference space. A benefit of our approach is that the spatial effects with the NCL are more tractable than a multivariate probit.  Although the structure and execution of \cite{dotson2018} is different than ours it is shares the basic idea that the covariance can be populated with functions that depend upon similarity of products. However, our similarity is based upon a physical space and not an attribute space.


\subsection*{Model Specification} \label{sec:model_spec}
We now specify the NCL model that captures the effects of spatial competition within our choice model. Again, our motivation is that products that reside next to each other may have greater influence on the consumer choice than those that are farther away.  
Our NCL model allows for such possibilities, but it does not impose spatial competition.\footnote{We omit an outside good since the category we analyze is mature and stable and we do not expect expansion or contraction in the category. The downside of this assumption is that our model must be interpreted conditional upon overall demand. Therefore, we cannot capture category expansion associated with product designs meant to encourage category expansion. This omission of the outside good is not a theoretical limitation, but a pragmatic one to simplify the analysis.} We define the utility of consumer $i$ from product $j$ in store $s$ during period $t$ as:

\begin{equation}
\label{implicitutilitywithattractiveness}
    u_{ijst} := 
    \pmb{x}_j^\intercal\pmb{\beta}^c - \beta^p_{ist}p_{jst} + \sum_{g=1...G} \mathpzc{a}_{g}I_g(\lambda^{d}_j) + \xi_{jst} + \epsilon_{ijst}
    = \nu_{ijst} + \epsilon_{ijst}
\end{equation}

where $\nu_{ijst}= \pmb{x}_j^\intercal\pmb{\beta}^c - \beta^p_{is}p_{jst} + \sum_{g=1...G} \mathpzc{a}_{g}I_g(\lambda^{d}_j) + \xi_{jst}$, $\pmb{x}_j$ is a vector of attributes associated with the product like its brand or size, $p_{jst}$ is the retail shelf price, $\mathpzc{a}_{g}$ represents the value of a fixed shelf  positional effect for location $g$ given a shelf partition structure and $I_g(\lambda^{d}_j)$ is an indicator variable for whether the center of product $j$'s facing is located within partition $g$ given shelf design $d$. $\xi_{jst}$ represents the unobserved product characteristics which vary across stores and time but are common to all customers and $\epsilon_{ijst}$ is the idiosyncratic error term. Notice the utility depends both upon the product attributes and shelf design ($\lambda$) through the shelf attractiveness terms ($\mathpzc{a}$). Each of these model elements: shelf design, attractiveness, and error distribution are discussed below.

\textbf{Shelf Design:} \label{sec:shelf_design}
The retailer's shelf planning decision can be thought of as a two-dimensional arrangement of products on a shelf. The challenge of shelf planning is the associated combinatorial explosion of potential designs. We define the \textbf{product set} of $J$ products as $\mathcal{J} := \{1,2,\cdots,J\}$. For example, suppose we have an example with five products, $J = 5$ products. Each product is indexed from $1$ to $5$. Therefore, the set of all products is $\mathcal{J} = \{1,2,3,4,5\}$. Let $\mathcal{L}$ denote the set of all possible product \textbf{shelf display locations}. These locations provide the potential coordinates or shelf position that each product may take.\footnote{These locations could either be continuous or discrete. Conceptually, it is probably easiest to think of discrete locations, but since planograms tend to correspond with physical shelves and products have varying dimensions, continuous locations are needed.} For example, suppose we have a shelf with a height of $15$ units and a width of $22$ units\footnote{Unit could be any standardized measurement such as feet or meters, or perhaps a common product width for the category. In this example, a unit represents six inches.} The set of all possible locations is $\mathcal{L} = [0,15]\times [0,22] \subset \mathbb{R}^2$. 



The arrangement of products on the shelf is determined by a \textbf{shelf design} function $\Lambda: \mathcal{J} \rightarrow \mathcal{L}$. This yields a complete set of mappings that determine a shelf design: $\Lambda := \{ \lambda_1, \lambda_2, \cdots, \lambda_J \}$. Each $\lambda_j$ refers to the location of product $j$. Products are often stacked on top of one another and blocked horizontally in a \textbf{product facing}. For simplicity, we always treat product facings as a unit or the area occupied by all items for product $j$. Therefore, $\lambda_j$ designates the centerpoint of the facing product $j$ on the shelf\footnote{We make a simplifying assumption that the depth of the shelf design does not enter into consumer decision making, although it could be relevant to the retailer in stocking the shelf. Also, we assume in our model that products are shelved contiguously or, equivalently, that a product will not be displayed in more than one product facing.}.


We denote the set of all \textbf{potential shelf designs} by $\mathcal{S}$. Our given shelf design is one instance of many potential shelf designs\footnote{Neither $\mathcal{L}$ nor $\mathcal{S}$ are assumed to be finite in general.}, $\Lambda \in \mathcal{S}$. For illustration purposes, we plot two shelf designs in Figure \ref{fig:D1D2}. We designate ``Design 1'' as $\Lambda^{(1)}=\{ \lambda_1^{(1)}, \cdots, \lambda_J^{(1)} \}$, and ``Design 2'' as $\Lambda^{(2)}=\{ \lambda_1^{(2)}, \cdots, \lambda_J^{(2)} \}$, where the superscript in parentheses refers to the specific shelf design\footnote{The specific values of the five products for the first shelf are $\lambda_{1}^{(1)} = (4,11)$, $\lambda_{2}^{(1)} = (10,11)$, $\lambda_{3}^{(1)} = (17,11)$, $\lambda_{4}^{(1)} = (8,4)$, and $\lambda_{5}^{(1)} = (18,4)$; and for the second shelf are $\lambda_{1}^{(2)} = (9,4)$, $\lambda_{2}^{(2)} = (4,11)$, $\lambda_{3}^{(2)} = (17,4)$, $\lambda_{4}^{(2)} = (14,11)$, and $\lambda_{5}^{(2)} = (3,4)$}.

\begin{figure}[H]
\centering
\input{Figures/figure_shelfdesign_a.tikz}
\caption{\footnotesize{Two possible shelf designs displaying products $\mathcal{J} = \{1,2,3,4,5\}$. The location of the centerpoint of each product facing, $\lambda_j^{(1)}$ and $\lambda_j^{(2)}$, is denoted by black dot labeled by product $j$.}}
\label{fig:D1D2}
\end{figure}

\textbf{Attractiveness:} Conceptually, the attractiveness term is a simple fixed effect that captures the corresponding value associated with the absolute shelf location where product $j$ resides.  This term is independent of shelf designs but it does rely on the partitioning of the shelf (e.g. how fine a shelf grid).\footnote{An important consideration in partitioning the shelf design is the number and location of the partitions} One could imagine a partitioning structure in which every product is assigned to its own partition. The challenge of such a granular partitioning structure is that it would take a great deal of variation in shelf design to measure the attractiveness of each partition. The other extreme is to simply have one partition for the entire shelf, but then this would negate the value of the partitions. Clearly, falling in between these two extremes is important. Our suggested heuristic is to define partitions as a grid made up of horizontal and vertical partitions. Our suggestion is to set the horizontal partitions to equal the number of shelves. The vertical partitions could be chosen to equal four to capture left, left center, right center, and right locations. For our paper towel example presented in Figure \ref{fig:D1D2withA}, it has two shelves with three vertical regions (left, center, and right) which yields 6 partitions. Alternatively if a finer measure of partition attractiveness is desired then the vertical partitions could be set to a size of roughly two or three product widths.  
These partitions are used to represent areas of interest such as eye-level, shelf header, and top or bottom shelves. This provides a convenient method for introducing positional effects into a shelf design. 
According to ``Design 1'' in Figure \ref{fig:D1D2withA}, product 1 is assigned to partition 1 and this product 1 has an attractiveness score of ${a_1}$. Alternatively if we follow ``Design 2'' then product 1 is assigned to partition 5, and has an attractiveness score of $a_5$.

\begin{figure}[ht]
\centering
\input{Figures/figure_shelfdesign_b.tikz}
\caption{\footnotesize{The previous shelf designs 1 and 2 are divided into 6 partitions $\mathcal{P} = \{\mathpzc{p}_1, \mathpzc{p}_2, \mathpzc{p}_3, \mathpzc{p}_4, \mathpzc{p}_5, \mathpzc{p}_6\}$. The partition is labeled in blue. Each partition has the associated attractiveness scores $\{a_1, a_2, a_3, a_4, a_5, a_6\}$.}}
\label{fig:D1D2withA}
\end{figure}

\textbf{Error Distribution:} We further capture the impact of shelf design and its relative position effects through the error correlation structure using the GEV model (\cite{1978modeling}). The cumulative distribution function associated with the multivariate GEV distribution is:

\begin{equation}\label{gevdistr}
    F(\epsilon_{i1st},\epsilon_{i2st},\cdots, \epsilon_{iJst}) := 
    \exp\left\lbrace
    -G\left({\rm e}^{-\epsilon_{i1st}}, {\rm e}^{-\epsilon_{i2st}},\cdots, {\rm e}^{-\epsilon_{iJst}} \right)
    \right\rbrace.
\end{equation} 
$G(\cdot)$ is referred to as the generator function and it permits correlated error structures. Notice that the MNL occurs as a special case when $G\left({\rm e}^{-\epsilon_{i1st}}, {\rm e}^{-\epsilon_{i2st}},\cdots, {\rm e}^{-\epsilon_{iJst}} \right)=\sum_{j=1}^{J}\epsilon_{ijst}$. Additionally, the nested logit model also occurs as another special case of the GEV (\cite{1978modeling}). In this paper McFadden demonstrated that if $G$ is non-negative, homogeneous of degree one and $y_j\geq 0$ $\forall j$, then it defines a probabilistic choice model consistent with random utility maximization (RUM) theory. Our NCL model uses the generator function:\footnote{in the empirical application below we set $\rho_{ij} := \rho$ for parsimony.}
\begin{equation}
\label{gev_gen}
    G(y_1,y_2,...,y_J)=\sum_{j=1}^{J-1} \sum_{j'=j+1}^{J} [(\alpha_{j,jj'}y_j)^{1/\rho_{jj'}}+(\alpha_{j',jj'}y_{j'})^{1/\rho_{jj'}}]^{\rho_{jj'}}
\end{equation}

where $y_j = exp(\nu_{ijst} + \epsilon_{ijst})$, $\nu_{ijst}$ is the deterministic part of the utility function (\ref{implicitutilitywithattractiveness}), $\alpha_{j,jj'}$ are the \emph{allocation parameters} and $\rho_{jj'}$ is the \emph{dissimilarity parameter}, both of which we discuss next.

\subsection*{NCL Choice Model:}

The maximization of random utility yields the choice probabilities for the NCL model:


\begin{equation}
\label{NCLmodel}
    P_{ist}^{NCL}(j)
    = \sum_{j' \neq j} P_{ist}(j|jj')P_{ist}(jj')
\end{equation}
where 
\begin{equation}
\label{NCLmodelC}
    P_{ist}(j|jj') = \frac{(\alpha_{j,jj'}e^{\nu_{ijst}})^{1/\rho_{jj'}}}
    {(\alpha_{j,jj'}e^{\nu_{ijst}})^{1/\rho_{jj'}} + (\alpha_{j',jj'}e^{\nu_{ij'st}})^{1/\rho_{jj'}}}
\end{equation}
represents the probability of choosing the product $j$ from the pair-nest $(j,j')$ and 

\begin{equation}
\label{NCLmodelM}
   P_{ist}(jj') = 
    \frac{\left((\alpha_{j,jj'}e^{\nu_{ijst}})^{1/\rho_{jj'}} + (\alpha_{j',jj'}e^{\nu_{ij'st}})^{1/\rho_{jj'}}\right)^{\rho_{jj'}}}
    {\sum_{k=1}^{J-1}\sum_{l=k+1}^J\left((\alpha_{k,kl}e^{\nu_{ikst}})^{1/\rho_{kl}} + (\alpha_{l,kl}e^{\nu_{ilst}})^{1/\rho_{kl}}\right)^{\rho_{kl}} }
\end{equation}

represents the probability of choosing the nest $(j,j')$ from every other possible pair-nest.  

An important aspect of this model, is that it has an analytical expression to quantify the degree of competition product $j$ faces in a given design, given by: $\sum_{j'}P_{j|jj'}$. Analogously, we can identify the product $j$ visibility in a given design as: $\sum_{j'}P_{jj'}$. These quantities move in opposing directions when product $j$ changes location, and none of these effects dominates in every situation, thus finding good designs becomes a very difficult task.

The motivation for this structure is to introduce the influence of a product to another based on their relative position in the shelf. Consider a subset or nest that includes only two products, we denote the pair-nest $(j,j')$ where $j, j' \in \mathcal{J}$. We assume consumers first select a pair-nest and then a product within the nest where the decision of selecting the nest pair $(j,j')$ is influenced by the spatial location of $j$ and $j'$ on the display, the spatial proximity between $j$ and $j'$ on the display, and their combined utilities. The choice between $j$ and $j'$ within the nest is modeled as a weighted MNL, where the weights will also include their relative proximity and indirectly the proximity of other products in the neighborhood. It is this feature that allows us to incorporate the attractiveness of the shelf location, for example the product is at eye-level, as well as the proximity and quality of competitors.

\textbf{The dissimilarity parameter $\rho_{jj'}$:} The correlation across nests is captured by the dissimilarity parameter $\rho_{jj'}$ which takes values in the interval $(0,1]$. If $\rho_{jj'} = 1$ then the pair-nest of products $j$ and $j'$ is completely dissimilar to other nests. On the other hand, as $\rho_{jj'} \rightarrow 0^+$ the pair-nest of products $j$ and $j'$ becomes more correlated with other pair-nests in the display which results in increased competition.

\textbf{Allocation parameters:} Following \cite{wen2001generalized} setup of the PGNL, we define the \emph{allocation parameters} $\alpha_{j,jj'}$ for any products $j,j' \in \mathcal{J}$ to be the fraction of a product $j$ in shelf design $\Lambda$ to be allocated to the pair-nest comprised by products $j$ and $j'$.  However, instead of estimating these parameters as in \cite{wen2001generalized}, we impose a functional structure based upon the proximity function to capture the correlation between products.

The allocation parameters are given by:
\begin{equation}
    \alpha_{j,jj'} := \frac{f_{jj'}}{\sum_{k \in \mathcal{J}}f_{jk}}.
\end{equation} 

and take larger values for products that are closer to one another. We observe that $\alpha_{j,jj'} \geq 0$ and $\sum_{j'\in \mathcal{J}}\alpha_{j,jj'} = 1$. Larger allocation parameters correspond to greater correlation between the utility of $j$ and $j'$. In turn this means greater competition between the pair.

Notice that our allocation parameters, $\alpha_{j,jj'}$, are dependent upon both the shelf design ($\Lambda$) and the proximity function ($f$).\footnote{If we would like to emphasize that we are working with the NCL model with a specific proximity function $f$ then we denote our model as $\text{NCL}^f$. We suppress adding superscripts to simplify our notation, unless it is important to stress that we are comparing between shelf designs or proximity functions.} 

\textbf{Proximity function $f$: } We introduce the proximity function $f: \mathcal{J}\times \mathcal{J}\rightarrow \mathbb{R}_{\geq 0}$ to measure the closeness between two products. Proximity is a function of the distance between the centerpoints of two products. Notice that the proximity function depends upon the shelf design $\Lambda$, but we suppress this in our notation for clarity. Formally, the proximity function needs to satisfy two properties: 

\begin{enumerate}
    \item $f_{kj} \geq f_{kj'}$ if $\lambda_k$ is closer to $\lambda_j$ than $\lambda_{j'}$. The proximity function is increasing in the distance between any two products $j$ and $j'$. By convention, we set the distance between product $j$ and itself to zero, $f_{jj} := 0$. 
    \item $f$ is symmetric: $f_{jj'} = f_{j'j}$. In other words the distance between $j$ and $j'$ is the same as the distance between $j'$ and $j$.
\end{enumerate}

We introduce two potential proximity functions used in our analysis: exponential, and inverse. Each function satisfies our two properties. Let $d_{jj'}$ be the Euclidean distance\footnote{We employ Euclidean distance for convenience of interpretation, but other distance functions may be used.} between locations $j$ and $j'$, and $\gamma$ be a scale parameter that is specific to each form.

\textit{Exponential proximity function.} In this case proximity follows an exponential function which decreases with distance: $f^{exp}_{jj'}:=e^{-\gamma d_{jj'}}$

\textit{Inverse proximity function.} We may wish a distance function that decays more quickly than the exponential proximity function, for which we propose an inverse power of distance: $f^{inv}_{jj'}:=\frac{1}{d_{jj'}^\gamma}$


Recall that the distances between locations are assumed known, but $\gamma$ is to be estimated in these first two functions.

\textbf{Demand influenced by proximity to competitors:} The desired characteristic of the model is that when two products are very close, they become strong substitutes of each other, as they get compared more often. This effect can be achieved by allowing shelf position to induce correlation in their demands. Specifically, to allow the demand of those two products to be highly correlated when products are close to each other, and low correlation otherwise. For simplicity, we assume that the influence of the shelf design is only a function of the separation between products and be equal for every pair of products ($\rho_{ij}=\rho$), otherwise the number of parameters would grow proportional to the squared of the number of products. A natural modeling choice is an extension of the pairwise nested logit model, because it allows for a full differentiated correlation structure while still retaining a tractable analytical form. This analytical form not only simplifies the estimation, but it also allows for the performing of analytical calculations on the choice probabilities to generate recommendations, and enables the use of optimization tools to find optimal shelf designs.

A disadvantage of this modeling choice is that, from equations \ref{NCLmodel}, \ref{NCLmodelC}, and \ref{NCLmodelM}, it is hard to visualize how demand correlation is induced by the proximity of competitors. When the dissimilarity parameter $\rho$ is equal to one (when no shelf induced correlation exists) the choice probabilities reduced to MNL (since $\sum_{j'\in \mathcal{J}}\alpha_{j,jj'} = 1$). In general when $0 < \rho < 1$, the correlation between two products cannot be written in closed form. However, the correlation can be computed using numerical integration (details in Appendix \ref{ap:correlation}). 

Table \ref{num_correl} presents the correlation between two identical products (to isolate the effect of proximity on their demand correlation), when their separation changes as a function of the dissimilarity parameter $\rho$ and distance parameter $\gamma$ using the exponential proximity function. In this table, we can observe that, as the dissimilarity parameter $\rho$ approaches 1, the demand correlation induced by the proximity between the two products approaches zero, regardless of their separation. In contrast, when the parameter gamma increases the correlation between the products also increases. This effect is stronger when the products are closer in proximity. More importantly, when $\gamma$ is closer to zero, the correlational effect decays slowly with separation producing a flat effect across the shelf. Conversely, when $\gamma$ is large, the display correlational effect matters only when the products are very close and it fades quickly away as separation increases.

The parameter $\rho$ and $\gamma$ together determine the overall degree of correlation induced by the display among products, but only $\gamma$ measures the decay of correlation with distance. Thus, differences in demand when products are separated by different distances allows the identification of these parameters separately.

\begin{table}[ht]
\caption{\footnotesize{Correlation between two products $i$ and $j$ as a function of the model parameters and their separation.}}
\singlespacing
\resizebox{\columnwidth}{!} {
\begin{tabular}{ c c c c c c c }
 Dissimilarity Parameter  & Distance Parameter in exponential function  & \multicolumn{5} {c@{}} {\# of Products between products $i$ and $j$} \\
 \cline{3-7}
  $\rho$ & $\gamma$ & adjacent & 1 & 2 & 3 & 4 \\  
  \hline
 0.2   &  0.1  & 0.067 &  0.052  & 0.039  & 0.030  & 0.024 \\
       &  0.2  & 0.157 &  0.097  & 0.053  & 0.032  & 0.020 \\
       &  0.3  & 0.296 &  0.156  & 0.059  & 0.028  & 0.015 \\
       &  0.4  & 0.451 &  0.222  & 0.057  & 0.022  & 0.010 \\
       &  0.5  & 0.582 &  0.289  & 0.048  & 0.016  & 0.006 \\
 \\
 0.4   &  0.1  & 0.058 &  0.045  & 0.034  & 0.026  & 0.021 \\
       &  0.2  & 0.137 &  0.084  & 0.046  & 0.028  & 0.018 \\
       &  0.3  & 0.257 &  0.136  & 0.052  & 0.025  & 0.013 \\
       &  0.4  & 0.391 &  0.193  & 0.050  & 0.020  & 0.009 \\
       &  0.5  & 0.502 &  0.252  & 0.043  & 0.014  & 0.005 \\
 \\
 0.6   &  0.1  & 0.044 &  0.034  & 0.026  & 0.02   & 0.016 \\
       &  0.2  & 0.103 &  0.064  & 0.035  & 0.021  & 0.013 \\
       &  0.3  & 0.192 &  0.102  & 0.039  & 0.019  & 0.010 \\
       &  0.4  & 0.290 &  0.145  & 0.038  & 0.015  & 0.006 \\
       &  0.5  & 0.370 &  0.188  & 0.033  & 0.011  & 0.004 \\
 \\
 0.8   &  0.1  & 0.024 &  0.019  & 0.014  & 0.011  & 0.009 \\
       &  0.2  & 0.056 &  0.035  & 0.019  & 0.012  & 0.007 \\
       &  0.3  & 0.104 &  0.056  & 0.022  & 0.010  & 0.005 \\
       &  0.4  & 0.155 &  0.079  & 0.021  & 0.008  & 0.004 \\
       &  0.5  & 0.197 &  0.102  & 0.019  & 0.006  & 0.002 \\
\\    
 1.0   &  0.1  & 0.000 &  0.000  & 0.000  & 0.000  & 0.000 \\
       &  0.2  & 0.000 &  0.000  & 0.000  & 0.000  & 0.000 \\
       &  0.3  & 0.000 &  0.000  & 0.000  & 0.000  & 0.000 \\
       &  0.4  & 0.000 &  0.000  & 0.000  & 0.000  & 0.000 \\
       &  0.5  & 0.000 &  0.000  & 0.000  & 0.000  & 0.000 \\

\label{num_correl}
\end{tabular}
}
\end{table}

\subsection*{Alternative Models} \label{sec:model_choice}

In addition to the above NCL model, we present several other models for comparison.  Below we present each  alternative. In the empirical results section, we will also discuss parameter estimates and model fit relative to our NCL model.


\textbf{Model specifications with and without fixed attractiveness.} We refer to the model (Equation \ref{NCLmodel}) as the NCL model when we would like to emphasize that this model includes fixed attractiveness effects. Alternatively, we consider the model without fixed attractiveness effects which is defined when the fixed attractiveness effects vanish ($\mathpzc{a}_g=0$ for all $g$) as:
\begin{equation}\label{implicitutilitywithoutattractiveness}
u_{ijst} := \pmb{x}_j^\intercal\pmb{\beta}^c - \beta^p_{is}p_{jst} + \xi_{jst} + \epsilon_{ijst} = \nu_{ijst} + \epsilon_{ijst}.
\end{equation}
We denote this model as the NCL$^\dagger$ model. For clarity we are introducing the $\dagger$ superscript to denote a variation on the model that does not have fixed attractiveness effects. Therefore NCL and NCL$^\dagger$ represent an NCL model with and without fixed attractiveness effects, respectively.

\textbf{Homogeneous NCL Model.} The homogeneous model assumes all consumers have the same response (or equivalently there is only one consumer per store), and we can drop the $i$ subscript. Additionally, we redefine the deterministic portion of utility as $\nu_{jst}=\pmb{x}_j^\intercal\pmb{\beta}^c - \beta^p_{s}p_{jst} + + \sum_{g=1...G} \mathpzc{a}_{g}I_g(\lambda_j)+\xi_{jst}$, and total utility to be $u_{jst} = \nu_{jst} + \epsilon_{jst}$. We define this as the homogeneous case of the NCL model, which we designate as the HNCL model. 
\begin{align*}
    P^{HNCL}_{st}(j) &= \sum_{j'\neq j}P_{st}(j|jj')P_{st}(jj')\\
    &= \sum_{j\neq j'}\frac{(\alpha_{j,jj'}e^{\nu_{jst}})^{1/\rho_{jj'}}}{(\alpha_{j,jj'}e^{\nu_{jst}})^{1/\rho_{jj'}} + (\alpha_{j,jj'}e^{\nu_{jst}})^{1/\rho_{jj'}}}\frac{\left((\alpha_{j,jj'}e^{\nu_{jst}})^{1/\rho_{jj'}} + (\alpha_{j',jj'}e^{\nu_{j'st}})^{1/\rho_{jj'}}\right)^{\rho_{jj'}}}{\sum_{k=1}^{J-1}\sum_{l=k+1}^J\left((\alpha_{k,kl}e^{\nu_{kst}})^{1/\rho_{jj'}} + (\alpha_{l,kl}e^{\nu_{lst}})^{1/\rho_{jj'}}\right)^{\rho_{jj'}}}.
\end{align*}

\textbf{Nested Logit Model.}
The choice probabilities for the nested logit model (NL) are given by
\begin{equation*}
    P^{NL}_{st}(j) 
    = P_{st}(j|{\lbrace \sigma_j \rbrace})P_{st}({\lbrace \sigma_j \rbrace}) 
    = \frac{e^{\nu_{jst}/\rho_{jj'}}}{\sum_{k \in \lbrace \sigma_j \rbrace}e^{\nu_{kst}/\rho_{jj'}}}\frac{\left(\sum_{k \in \lbrace \sigma_j \rbrace}e^{\nu_{kst}/\rho_{jj'}}\right)^{\rho_{jj'}}}{\sum_{k=1}^{|\mathcal{P}|}\left(\sum_{l \in \lbrace \sigma_k \rbrace}e^{\nu_{lst}/\rho_{jj'}}\right)^{\rho_{jj'}}},
\end{equation*}
where $\lbrace \sigma_k \rbrace$ refers to the set of products that have been assigned to partition $k$.
 Each product $j$ belongs to exactly one partition ($\sigma_j$) with products nested by partition, so the allocation parameters are given by $\alpha_{j,q} = 1$ if $j \in \lbrace \sigma_j \rbrace$ and $0$ otherwise. As before, we refer to the model without the fixed attractiveness effects as NL$^\dagger$ model.

\textbf{Comparison with the MNL Model.} The position-dependent variant of the MNL model is  similar to \cite{dreze1994shelf}, in that we introduce the location attractiveness into the utility by including measures of the product's horizontal and vertical coordinates. For each product $j \in \mathcal{J}$ we denote by $\pmb{x}_j$ a vector representing the attributes of product $j$. Assuming a linear form, the implicit utility $u_{jst}$ of product $j$ is defined as in (\ref{implicitutilitywithattractiveness}) but assumes $\epsilon_{jst}$ distributed iid with type I Extreme Value. The probability of selecting the product $j$ from the display of a set of product $\mathcal{J}$  follows the logistic form 
\begin{equation}\label{mnlmodel}
    P_{st}^{MNL}(j) = \frac{e^{\nu_{jst}}}{\sum_{k = 1}^Je^{\nu_{kst}}}
\end{equation}
As before we refer to the MNL with positional effects as the MNL model, and the model without the fixed shelf position effects as MNL$^\dagger$.

Since the random term is iid, the correlation between alternatives is then only driven by observed attributes.
Conversely, the NCL model incorporates the product display design thus inducing correlation between neighboring products based on their relative position in the shelf. This effect has been neglected in traditional choice models in marketing. 

When compared to the MNL where only products with desirable attributes will receive attention from the consumers, the NCL model provides us with some additional insights. When a product is considered, consumers are also likely to consider a group of products in the close neighborhood as well. Therefore, a product not only gains attention when placed in an attractive location and sufficiently far from other desirable products (competition effect), but it also benefits from other products attractiveness in its neighborhood (visibility effect). Being placed near attractive products has the advantage of higher attention and the disadvantage of being compared more often with strong competitors, so the optimal location is the result of balancing these two effects.

By adjusting the dissimilarity parameter $\rho$ appropriately, the HNCL and HNCL$^\dagger$ models can be shown to reduce to MNL and MNL$^\dagger$ models respectively, $P^{HNCL}_{jst}|_{\rho = 1} = P^{MNL}_{jst}, \qquad$ and $\qquad P^{HNCL^\dagger}_{jst}|_{\rho = 1} = P^{MNL^\dagger}_{jst}$. This is unsurprising when $\rho = 1$, as any pair of products $j, k \in \mathcal{J}$ becomes completely uncorrelated and we have the usual independent error structure of the MNL.

On the other hand, as $\rho \rightarrow 0^+$, product pair $j,k \in \mathcal{J}$ becomes more correlated. In this limit HNCL or HNCL$^\dagger$ model will not reduce to MNL or MNL$^\dagger$ model as the final result still depends on the relative locations between products. However, we can still say that NCL and NCL$^\dagger$ model reduce to the MNL$^\dagger$ model for choosing a pair-nest from the set of all possible nests. Suppose that we fix the shelf design $\Lambda$ then the HNCL or HNCL$^\dagger$  with utilities $\nu_{jst}$ is equivalent to the MNL model with utilities $\tilde{\nu}_{jst} := \log \left(\sum_{k \in \mathcal{J}_j }\alpha_{j,jk}e^{\nu_{jst}}\right)$, where 
\begin{equation*}
    \mathcal{J}_j := \{ k \in \mathcal{J}\ |\ \alpha_{j,jk}e^{\nu_{jst}} > \alpha_{k,jk}e^{\nu_{kst}}\}.
\end{equation*}

\section*{An Empirical Analysis of Shelf Design for Paper Towels}\label{sec:empirical}

In this section we conduct an empirical analysis of the paper towel dataset from Dominick's Finer Foods. A key advantage of this dataset is the presence of exogenous shelf experiments which introduces variation in shelf designs. A common problem with sales datasets is that there may be no natural variation in shelf design. Many retailers employ a uniform shelf design across all stores or the variation is endogenous. Additionally, there is a natural confound between the shelf design and the store design. For example, large stores may employ expanded displays or stores in urban neighborhoods may employ reduced sets. Therefore, the natural variation present may be insufficient or confounded with other factors. However, in our dataset we can rely upon artificial variation in shelf design to estimate our model. We estimate the NCL model as well as alternative benchmark models to compare the fit and predictive accuracy. Furthermore, we conduct a counterfactual analysis to find whether an even more profitable design can be created than those in our dataset.

Before we begin to discuss how we estimate the heterogeneous model, we must first discuss how we partition the store shelf and present several assumptions to ensure the model is tractable. The partitioning of the store shelf is relatively straight forward. The store shelves have three rows, high, middle and low, so we restrict our vertical partitions to these.  Next we assume that the store shelf is partitioned into 6 equally spaced horizontal area.  Thus, partitioning the store shelf into  18 distinct areas.\footnote{We discuss the robustness to this partitioning structure in the Robustness Check}  This partitioning of shelves is illustrated in Figure \ref{fig:shelfwithattr}.

Next, we discuss several important modeling assumptions for estimation. The first is that we fix $\gamma$ within the proximity function to be constant across all pairs $j$ and $j'$. This assumption greatly reduces the complexity of estimation by reducing the number of parameters with respect to GNL while still preserving the influence of shelf proximity on demand.  The next assumption is with respect to the dissimilarity parameter $\rho_{j,j'}$.  In our empirical analysis we assume the dissimilarity is constant across all product pairs ($\rho_{jj'}=\rho$ $\forall j,j'\in \mathcal{J}$).  Again this assumption greatly reduces the complexity of estimation by reducing the number of parameters to estimate while still preserving the influence of shelf proximity on demand. 
 
\subsection*{Estimation of Heterogeneous Model} \label{sec:heterogeneous} 

Each store in the data were recorded to have different customer demographic. For the purpose of our study let us consider the differences in consumer's average income $I_s$ across the stores. Let us recall from (\ref{implicitutilitywithattractiveness}) that the implicit utility of product $j \in \mathcal{J}$ for the individual $i$ at the store $s$ in the time frame $t$ is given by
\begin{equation}
    u_{ijst} := \pmb{x}_j^\intercal\pmb{\beta}^c - \beta^p_{is}p_{jst} +  \sum_{g=1...G} \mathpzc{a}_{g}I_g(\lambda^{d}_j) + \xi_{jst} + \epsilon_{ijst}
\end{equation}
where $p_{jst}$ is the price of product $i$, $\xi_{jst}$ represents the unobserved product characteristics for product $j$ at store $s$ in the time $t$ which is common to all customers, and $\epsilon_{ijst}$ is the idiosyncratic error term. For us, each time frame corresponds to the data collected in each week at each different store. The 
price sensitivity $\beta^p_{is}$ is dependent on income which varies across stores. The price coefficient is given by
    
    \begin{equation}
       \beta^p_{ist}=\beta^p + {\Pi}D_{is} + {L}v_i,
       \end{equation}

where ${\Pi}$ is the observed heterogeneous price effect due to demographic  $D_{is}$ and ${L}$  is unobserved heterogeneous component. The demographic factor is given by
\begin{equation}
    D_{is} := I_{s} + \omega_sw_i
\end{equation}
where $w_i \sim \text{ i.i.d. } \mathcal{N}(0,1)$, $I_{s}$ is the average income of customers of the store $s$, and $\omega_s$ is the standard deviation of $I_s$. This specification is feasible given access to the demographic information of consumers at each store. 

Finally, the probability of a randomly chosen individual $i$ choosing a product $j$ at the store $s$ in the time frame $t$ according to the model, $M \in \{MNL, MNL^{\dagger}, NCL^{exp}, NCL^{inv}, \cdots\}$, is given by the expected value of $P^M_{ist}(j)$ over the demographic distribution:

\begin{equation}
    P^M_{st}(j) := \int_{D_{is} \in \mathbb{R}}\int_{\pmb{v}_{i}\in\mathbb{R}^{J+1}}P^M_{ist}(j)dF(\pmb{v}_i)dF(D_{is})
\end{equation}

where $F(D_{is})$ and $F(\pmb{v}_{i})$ denotes the Normal cumulative distribution of $D_{is}$ and $\pmb{v}_i$ respectively. In practice, the integral above can be approximated using numerical integration by the random draw of $R$ individuals $i$ from the given distributions:

\begin{equation}
    P^{M}_{st}(j) \approx \frac{1}{R}\sum_{i = 1}^R P^{M}_{ist}(j).
\end{equation}

To estimate the heterogeneous model, we need to estimate both the homogeneous parameters $\pmb{\theta}_1 := (\beta^p, \pmb{\beta}^c, a_g, \gamma, \rho)$ and the new heterogeneous parameters $\pmb{\theta}_2 := ({\Pi}, {L})$. This can be done using the BLP technique following \cite{berry1995automobile} which we summarize as follows: 

\begin{enumerate}
    \item Let $S_{jst}$ be the observed market share of product $j$ at the store $s$ in the time frame $t$. Define the mean homogeneous implicit utility $\delta_{jst} := \delta_{jst}(\pmb{\theta}_2, \gamma, \rho)$ \footnote{$\delta_{jst}$ is a function of $\pmb{\theta}_2$ but it is also a function of $\gamma$ and $\rho$ when $M$ is one of the NCL models. We recall that $P^{NCL}_j$ as given in (\ref{NCLmodel}) depends on the dissimilarity parameter $\rho$ and its dependency on the scale parameter $\gamma$ follows from our choice of the proximity functions.} to be the function given by the solution of the equation:
    \begin{equation}
        S_{jst} = \frac{1}{R}\sum_{i=1}^R P^{M}_{ijst}\left(\left\{u_{ijst} = \delta_{jst} + (-p_{jst}, \pmb{x}_j^\intercal)\left({\Pi}D_{is} + {L}{v}_i\right)\right\}\right).
    \end{equation}
    In other words, we have $\xi_{jst} = \delta_{jst} - (\pmb{x}^\intercal_j\pmb{\beta}^c - \beta^p p_{jst} + \sum_{g=1...G} \mathpzc{a}_{g}I_g(\lambda^{d}_j))$.
    \item Choosing the instrumental variables $z_{jlst}, l = 1,\cdots, L$ for some $L\geq 1$. As instruments, we use average acquisition costs\footnote{$c_{jst} := \text{UnitsPurchased}_{jst}\cdot \text{WholeSalePrice}_{jst} + \left(\text{Inventory}_{jst} - \text{Sales}_{jst}\right)\cdot c_{js,t-1}$} as specified in \cite{peltzman2000prices}. In our model, we use $L=J$ and $z_{jlst} := c_{jst}\pmb{1}_{j=l}$, where $c_{jst}$ is the cost of product $j$ at the store $s$ in time frame $t$ and $\pmb{1}_{i=j}$ denotes the selection function which is $1$ if $i=j$ and $0$ otherwise.
    \item Solve for $\pmb{\theta}_1$ and $\pmb{\theta}_2$ such that $\sum_{j,s,t}z_{jlst}\xi_{jst} = 0$ $\forall l = 1,\cdots, L$, where the summation runs over all products $j \in \mathcal{J}$, all stores $s = 1,\cdots,S$ and time frames $t=1,\cdots,T$ where $S$ and $T$ are the number of stores and time frames respectively. Equivalently, we find $\pmb{\theta}_1$ and $\pmb{\theta}_2$ that minimize the following objective function:
    \begin{equation}
        Min_{\theta_1,\theta_2}  (\pmb{Z'\xi({\theta_1,\theta_2})})'\pmb{W}(\pmb{Z'\xi({\theta_1,\theta_2})})  
    \end{equation}
    where \pmb{W} is the weighting matrix.
\end{enumerate}

\subsection*{Estimation Results}\label{sec:homogeneous}
Below we present the empirical estimates for our NCL models as well as several benchmarks.  In order to estimate the NCL model we need to specify a specific form for $f_{ij}$. We use two different functions $f^{inv}_{\cdot,\cdot}$ and $ f^{exp}_{\cdot,\cdot}$ to estimate two versions of NCL model.  We also estimate those two models with and without the attractiveness score function, $NCL^{inv}$, $NCL^{inv\dagger}$ and $NCL^{exp}, NCL^{exp\dagger}$ respectively. 

In addition to the two versions of the NCL model, we fit the MNL model, MNL without location attractiveness (denoted as MNL$^\dagger$), and consider the NL and NL$^\dagger$ models.

\begin{table}[ht!]
\centering
\resizebox{0.6\textwidth}{!}{\begin{tabular}{lccccc}
    \toprule
                   & \textbf{RMSE}      & \textbf{RMSE}       &  \textbf{$\gamma$} & \textbf{$\rho$}   & \textbf{Price} \\
     \textbf{Model}& \textbf{In-Sample} & \textbf{Out-Sample} & (per hundred inches) &                   & \textbf{sensitivity} \\
    \hline
\multirow{2}{*}{$MNL$} & \multirow{2}{*}{\color[rgb]{0.4989122184160447,0.5010877815839553,0} 0.99\%} & \multirow{2}{*}{\color[rgb]{0.9900435551391445,0.009956444860855451,0} 1.57\%} & --- & --- & 3.44\\ 
 & &  & --- & --- & (0.02)\\ 
\multirow{2}{*}{$MNL^\dagger$} & \multirow{2}{*}{\color[rgb]{1.0,0.0,0} 1.99\%} & \multirow{2}{*}{\color[rgb]{1.0,0.0,0} 1.59\%} & --- & --- & 3.32\\ 
 & &  & --- & --- & (0.01)\\ 
 \multirow{2}{*}{$NL$} & \multirow{2}{*}{\color[rgb]{0.4994380881390911,0.5005619118609089,0} 0.99\%} & \multirow{2}{*}{\color[rgb]{0.9443379713837988,0.055662028616201176,0} 1.50\%} & --- & 0.96 & 3.43\\ 
 & &  & --- & (0.00) & (0.02)\\ 
\multirow{2}{*}{$NL^\dagger$} & \multirow{2}{*}{\color[rgb]{0.9306546529101654,0.06934534708983464,0} 1.85\%} & \multirow{2}{*}{\color[rgb]{0.899696040177643,0.10030395982235696,0} 1.43\%} & --- & 0.89 & 3.32\\ 
 & &  & --- & (0.01) & (0.02)\\ 
 \multirow{2}{*}{$NCL^{exp}$} & \multirow{2}{*}{\color[rgb]{0.0,1.0,0} 0.00\%} & \multirow{2}{*}{\color[rgb]{0.27907452712692565,0.7209254728730743,0} 0.44\%} & 0.27 & 0.25 & 2.73\\ 
 & &  & (0.22) & (0.01) & (0.06)\\ 
\multirow{2}{*}{$NCL^{exp\dagger}$} & \multirow{2}{*}{\color[rgb]{0.2638336099192982,0.7361663900807018,0} 0.53\%} & \multirow{2}{*}{\color[rgb]{0.0,1.0,0} 0.00\%} & 0.01 & 0.25 & 2.73\\ 
 & &  & (0.01) & (0.11) & (0.10)\\ 
 \multirow{2}{*}{$NCL^{inv}$} & \multirow{2}{*}{\color[rgb]{0.03271712703026738,0.9672828729697326,0} 0.07\%} & \multirow{2}{*}{\color[rgb]{0.27632460960421146,0.7236753903957885,0} 0.44\%} & 10.44 & 0.25 & 2.74\\ 
 & &  & (5.18) & (0.01) & (0.02)\\ 
\multirow{2}{*}{$NCL^{inv\dagger}$} & \multirow{2}{*}{\color[rgb]{0.26749297071611855,0.7325070292838815,0} 0.53\%} & \multirow{2}{*}{\color[rgb]{0.005600143001791685,0.9943998569982083,0} 0.01\%} & 0.00 & 0.25 & 2.73\\ 
 & &  & (0.00) & (0.01) & (0.02)\\ 
\bottomrule 
\end{tabular}} 
\caption{\footnotesize{Estimation with Paper Towel (PTW) Category.}}
\label{tab:PTW}
\end{table}

\begin{figure}[!ht]
    \centering
    \includegraphics[scale=0.25]{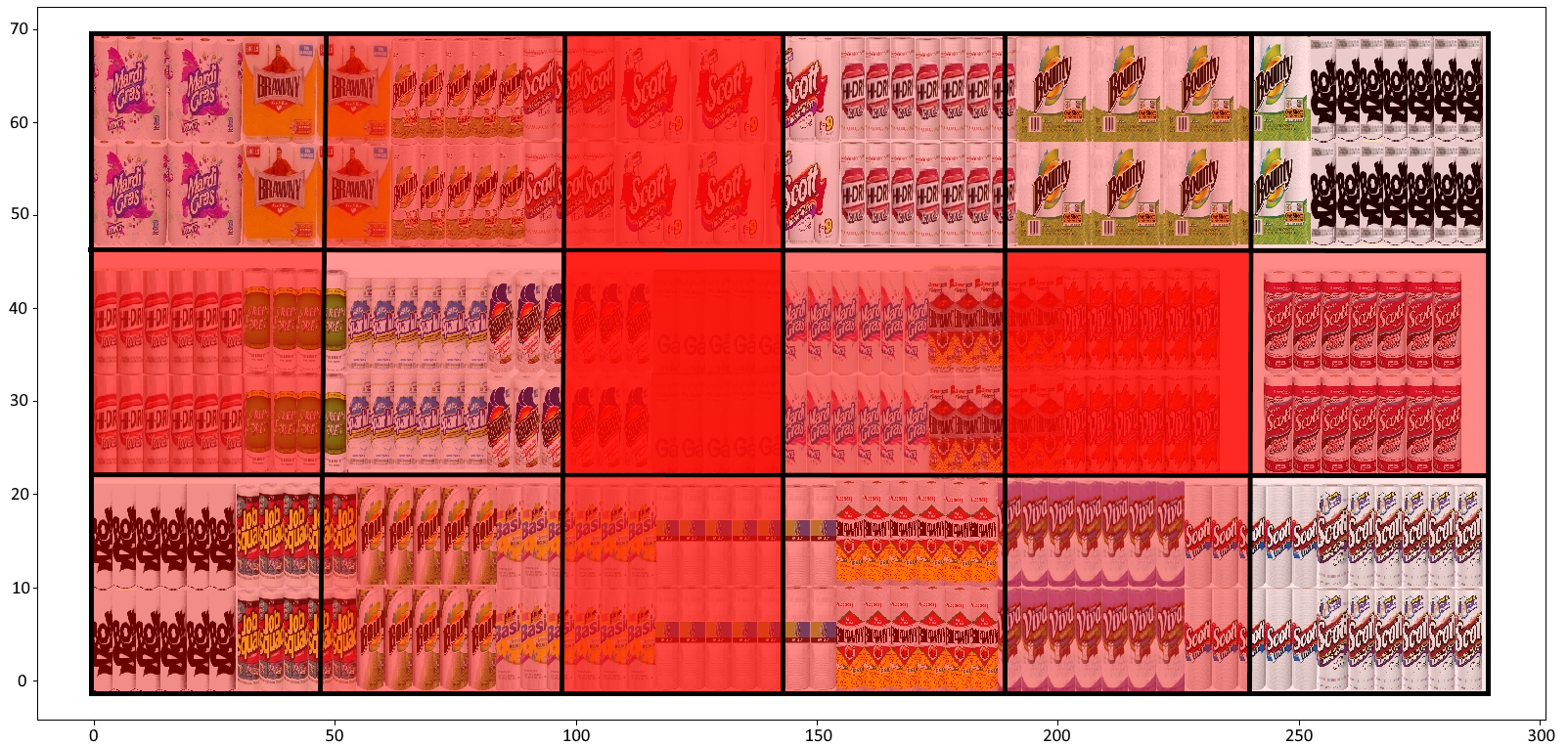}
    \caption{\footnotesize{Attractiveness score heat map on the shelf which is divided into 18 partitions. The higher intensity of the color red indicates the higher attractiveness score of the corresponding partition.}}
    \label{fig:shelfwithattr}
\end{figure}

\textbf{Results.} For the sake of completeness, we present the estimation results for the non-heterogeneous models in Table \ref{tab:PTW} and in Table  \ref{tab:PTW_BLP} for the models with consumer heterogeneity. The fixed shelf position effect ($a_g \forall g$) estimated from each relevant model is quite similar. Rather than listing all fixed effect point estimates for each corresponding shelf location, we illustrate the result from the $NCL^{exp}$ with a heat map in Figure \ref{fig:shelfwithattr} and determine that the center shelf regardless of row is an extremely attractive location.  Additionally, the right and left ends are superior locations as long as they are in the center row. 

Addressing the relative performance across all models requires that we calculate the \textit{out of sample} and within sample root means squared errors (RMSE). In practice, that entails holding out the control design group during estimation. In each calculation we set the structural error $\xi$ to zero. We observe the NCL model has a better fit in sample and predicts better than the benchmark models on the out-sample data. The parameter which captures the nest correlation, $\rho$, is significantly different from zero suggesting that the model differentiates from the MNL counterpart. This supports the conclusion that shelf interactions significantly contribute to improve model fit and predictions. 

Additionally, we present estimates of the heterogeneous parameter estimates $\Pi$ and $L$, where $\Pi$ is the estimate associated with the observed heterogeneity due to income and $L$ is the estimate of the standard deviation of the unobserved heterogeneity.  For each model we determine that there is significant and sizeable observed and unobserved heterogeneity in consumer preferences towards price.  The only exception is with the $NCL^{inv\dagger}$ model where observed heterogeneity is found to be insignificant at the 95\% confidence level.

\begin{table}[ht]
\centering
\resizebox{0.8\textwidth}{!}{\begin{tabular}{lccccccc}
    \toprule
                   & \textbf{RMSE}      & \textbf{RMSE}       &                   &                 & \textbf{Price} \\
     \textbf{Model}& \textbf{In-Sample} & \textbf{Out-Sample} & \textbf{$\gamma^*$} & \textbf{$\rho$} & \textbf{sensitivity} & $\Pi_0^*$ & $L_{0,0}^*$ \\
    \hline 
\multirow{2}{*}{$MNL$} & \multirow{2}{*}{\color[rgb]{0.6540344433129324,0.34596555668706763,0} 1.49\%} & \multirow{2}{*}{\color[rgb]{0.7425309086449727,0.25746909135502727,0} 2.06\%} & --- & --- & 3.22 & -7.61 & 8.83\\ 
 & &  & --- & --- & (0.00) & \small{(0.004)} & \small{(0.001)}\\ 
\multirow{2}{*}{$MNL^\dagger$} & \multirow{2}{*}{\color[rgb]{1.0,0.0,0} 2.28\%} & \multirow{2}{*}{\color[rgb]{1.0,0.0,0} 2.77\%} & --- & --- & 3.12 & -6.19 & -10.20\\ 
 & &  & --- & --- & (0.00) & \small{(0.003)} & \small{(0.002)}\\ 
  \multirow{2}{*}{$NL$} & \multirow{2}{*}{\color[rgb]{0.614338746025182,0.385661253974818,0} 1.40\%} & \multirow{2}{*}{\color[rgb]{0.6460793203429146,0.3539206796570854,0} 1.79\%} & --- & 0.96 & 3.22 & -7.22 & -2.38\\ 
 & &  & --- & (0.00) & (0.00) & \small{(0.002)} & \small{(0.000)}\\ 
\multirow{2}{*}{$NL^\dagger$} & \multirow{2}{*}{\color[rgb]{0.9187353532728061,0.08126464672719391,0} 2.10\%} & \multirow{2}{*}{\color[rgb]{0.9354247526569579,0.0645752473430421,0} 2.60\%} & --- & 0.89 & 3.15 & -4.96 & 9.03\\ 
 & &  & --- & (0.00) & (0.00) & \small{(0.003)} & \small{(0.001)}\\ 
 \multirow{2}{*}{$NCL^{exp}$} & \multirow{2}{*}{\color[rgb]{0.0011729408041854553,0.9988270591958145,0} 0.00\%} & \multirow{2}{*}{\color[rgb]{0.016329682402550765,0.9836703175974493,0} 0.05\%} & 0.26 & 0.25 & 2.70 & -0.68 & -2.78\\ 
 & &  & (0.05) & (0.00) & (0.00) & \small{(0.167)} & \small{(0.047)}\\ 
\multirow{2}{*}{$NCL^{exp\dagger}$} & \multirow{2}{*}{\color[rgb]{0.24770673041912605,0.752293269580874,0} 0.56\%} & \multirow{2}{*}{\color[rgb]{0.165485932544984,0.834514067455016,0} 0.46\%} & 0.05 & 0.24 & 2.70 & -0.63 & -3.05\\ 
 & &  & (0.06) & (0.00) & (0.00) & \small{(0.166)} & \small{(0.043)}\\ 
 \multirow{2}{*}{$NCL^{inv}$} & \multirow{2}{*}{\color[rgb]{0.06185378553015753,0.9381462144698425,0} 0.14\%} & \multirow{2}{*}{\color[rgb]{0.049221391872559576,0.9507786081274404,0} 0.14\%} & 12.55 & 0.25 & 2.72 & -0.38 & -2.94\\ 
 & &  & (0.04) & (0.00) & (0.00) & \small{(0.055)} & \small{(0.030)}\\ 
\multirow{2}{*}{$NCL^{inv\dagger}$} & \multirow{2}{*}{\color[rgb]{0.24699188082810244,0.7530081191718976,0} 0.56\%} & \multirow{2}{*}{\color[rgb]{0.16490927774825,0.83509072225175,0} 0.46\%} & 1.84 & 0.24 & 2.71 & -0.14 & 3.23\\ 
 & &  & (0.02) & (0.00) & (0.00) & \small{(0.043)} & \small{(0.019)}\\
\bottomrule 
\end{tabular}
\caption{\footnotesize{Estimation of heterogeneous model with Paper Towel (PTW) Category. Asterisks (*) indicate the values in the column are multiplied by 100 to ease the presentation. Product Fixed Effects are included}}
\label{tab:PTW_BLP}
}
\end{table}

Our estimation results in the full specification with exponential function $NCL^{exp}$ (table \ref{tab:PTW}) show a $\rho=0.25$, and $\gamma=0.27$ per hundred inches. The $\rho$ is significantly different from 1, which suggest a better fit of this model as compared to the traditional MNL, and $\gamma$ is positive and significantly different from zero, which suggest a significant decay in the demand correlation when separation increases. According to table \ref{num_correl}, these estimates suggest that when two products are very close, the proximity induced correlation is a little less than 0.3, while it is close to zero when they are separated by approximately 4 sets of products (90 inches of separation) or more.

Interestingly, when we consider the model without accounting for the shelf attractiveness $NCL^{exp\dagger}$, $\gamma$ estimate is very small, suggesting that in that specification the proximity is not relevant. The analysis with the inverse of distance function is analogous.

Finally, there is also a large change in the price coefficient of the NCL models relative to the MNL, MNL$^\dagger$, NL, and NL$^\dagger$. By not considering the interaction in the shelf design,  there appears to be a bias in the estimated price sensitivity. Incorrect price sensitivity would seriously hinder pricing and promotion decisions for both retailers and manufacturers. 

\subsection*{Robustness Check of Partitions} \label{partitions_check}

\begin{figure}
    \centering
    \includegraphics[width=0.6\textwidth]{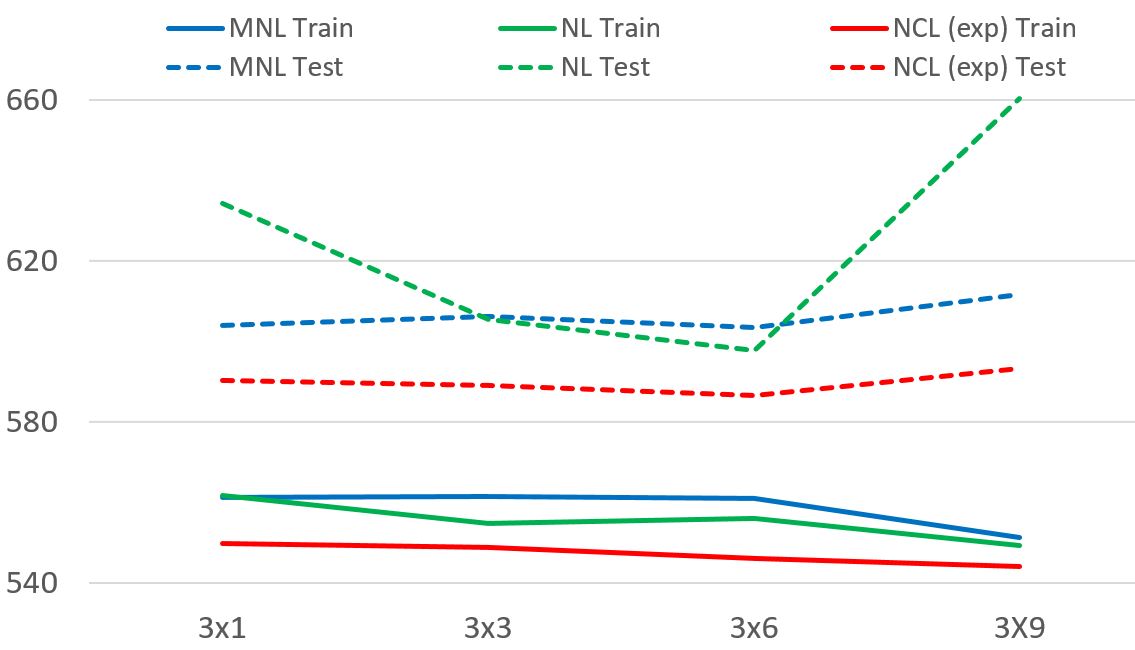}
    \caption{\footnotesize{Root Mean Square error of $MNL$, $NL$ and $NCL^{exp}$ models with the training data and testing data (Design 3). }}
    \label{fig:grid_check}
\end{figure}

The shelf fixed effects, reflected in the partition in our model, are not the key feature of the proposed model. However, choosing a wrong partition may affect the overall performance of the model. We test the performance of the model using different partition sizes. In our data it is natural to consider each row as a separate partition, since our shelf have three levels, so we varied how many horizontal partitions we include, for example, Figure \ref{fig:shelfwithattr} has six horizontal groups and three vertical, giving a shelf of 3x6 partitions. We examined partitions of 3x1, 3x3, 3x6 and 3x9 and present the error in the training and testing data in Figure \ref{fig:grid_check}.

As expected, when increasing the number of partitions, the RMSE goes down, however, when applying the model to new data, the error initially drops, but then goes up again. This suggests that the model with too many partitions tends to over fit.

We also test the model with partition of unequal sizes, for example a 3x3 partition with narrow left and right sides with a large center. This partition performed better than the 3x3 partition with equal size partition, but not better than the 3x6 partition.\footnote{After fitting the model, the model was used to predict the market shares of the products, then that vector was compared with real sales to compute RMSE.}

\subsection*{Identification}
    
We discuss the identification of the parameters by asking the question what variation in the data permits the estimation of each of the parameters. First, we discuss shelf attractiveness effects.  The identification of these parameters are relatively straightforward.  What is required is variation in the shelf design holding the partitions fixed across stores and time. Together, these variations along with variation in sales allows for the identification of these parameters.

Second is the identification of $\rho$.  It is important to highlight that as $\rho$ goes to 1 we converge to a multinomial logit model.  With $\rho$, we require variation in pair nest attractiveness levels. Thus, what pins down $\rho$ is the correlation between product sales and the pair nest attractiveness relative to other pair nests that include product $j$. In our data we observe variation in this pair attractiveness both when products are located in different locations in the shelf, and when the products themselves change their attractiveness with promotions or other price variation.

A third parameter of interest is $\gamma$, a measure that captures the importance of shelf spacing.  Generally speaking $\gamma$ is pinned down with co-variation in sales data across products within a pair nest when their separation varies across the designs, whereas $\rho$ leverages variation across pair nests. Of note is the importance that \textbf{variation in shelf design} and variation of observed and unobserved characteristics play in pinning down these important model parameters. Variation a focal product demand (sales) given variation in product attributes (e.g. price reduction or a change in product promotion) of products at  different distances from the focal product provides additional data variation to assist in identifying these important model parameters. For instance, the focal product demand may decrease from a price drop of a product nearby but yield no change in demand from a similar price drop of a product further from the focal item. Note that although $\rho$ and $\gamma$ together account for sales correlation of pair of products, only $\gamma$ allows for the decay in correlation when separation increases.

The last parameter of interest is price sensitivity.  Given that price is endogenous, we need to instrument to control for correlation between product price and the unobserved product characteristics. With product cost data available (an unusual opportunity for empirical work), we leverage this data by using the instrumental variables: $z_{jlst}$. For the $3\times 6$ shelf partitions, we may perform the following regression
\begin{equation*}
    p_{jst} = \pmb{x}_j^\intercal \pmb{\eta}^p + \pmb{z}_{jst}^\intercal \pmb{\eta}^z +  \sum_{g=1}^{17}I_g(\lambda^d_j)\eta^a_g + \tilde{\varepsilon}_{jst}
\end{equation*}
to find that the coefficients $\pmb{\eta}^p, \pmb{\eta}^z$, and $\pmb{\eta}^a$ are non-zero at the $99.9\%$ significant level with F-value $4.036\times 10^{4}$ or $p < 2.2\times 10^{-16}$.

\section*{Understanding Competition Induced by Shelf Design}  \label{sec:model_compete}

In this section we discuss properties of the NCL model as they relate to shelf competition. \cite{dreze1994shelf} show that there are better positions on a shelf. This research corroborates our intuition that certain physical locations like being at eye-level or the shelf-shoulder that a consumer sees first when entering the category are beneficial. However, these are fixed shelf effects. Our motivation for the NCL model is to allow for the potential that products may benefit from their position to desirable products in the category. 

Our conjecture is not that shelf design provides only a fixed effect but a relative effect that depends upon the attractiveness of the neighboring products. \cite{sayman2002positioning} show that store brands which imitate leading national brands benefit from being perceptually close. One way that retailers do this is through shelf placement. They present both a theoretical model and empirical evidence to support this argument. We believe that when consumers search for a leading national brand that other products might benefit from being nearby. Alternatively, spatial effects could work in the opposite direction in that being close to a dominating product both in terms of quality and price could draw buyers away from the nearby products to the dominating brand.

The NCL model permits a general framework in which to understand how shelf position permits a variety of spatial effects. The NCL model retains the advantages of the choice models while at the same time allowing flexibility in absolute and relative shelf position effects due to shelf design. In this section we consider several properties of the NCL model in greater depth to illustrate its ability to capture relative shelf effects. Specifically, in the next section we start to show how the relative attractiveness of a product is influenced by its position. Second, we continue in the following section to show that relative position can induce competition through the changes in the price elasticity matrix. In order to clearly illustrate these effects and provide the intuition in a simple manner, we present a stylized example that employs the relevant and important model estimates from above. In each case, we simplify the model with use of an example that allows for only a one dimensional linear shelf. 

\subsection*{Illustrating Relative Attractiveness of a Product Shelf Design}\label{threeproductexamplesection}

To understand the effect of moving a product in a shelf, let us consider a simple single row shelf but with the spacing of that shelf similar to what we see in the Dominick's setting. Suppose there are three products $\{L,X,R\}$ with utilities $\nu_L, \nu_X, \nu_R$. 
Using the parameters estimated and employed in the full model above (e.g. dissimilarity parameter $\rho = 0.25$). We assume the shelf is of length 300 inches with two low value products set at locations -25 and +25 inches. In this setting we analyze the effects of a third product that has either high or low value compared to the fixed products and vary its location along the shelf.  

The benefit of fixing the competing products is that we may isolate the impact of relative shelf position on the attractiveness of the target product $X$ exclusively as a function of its location. To understand the impact of varying the position of our target product $X$ relative to these other two products we focus on the impact of the probability of purchase of product $X$ denoted $P^{NCL}_X$. 

\begin{figure}
    \begin{subfigure}[t]{0.47\textwidth}
        \centering
        \includegraphics[width=0.94\textwidth]{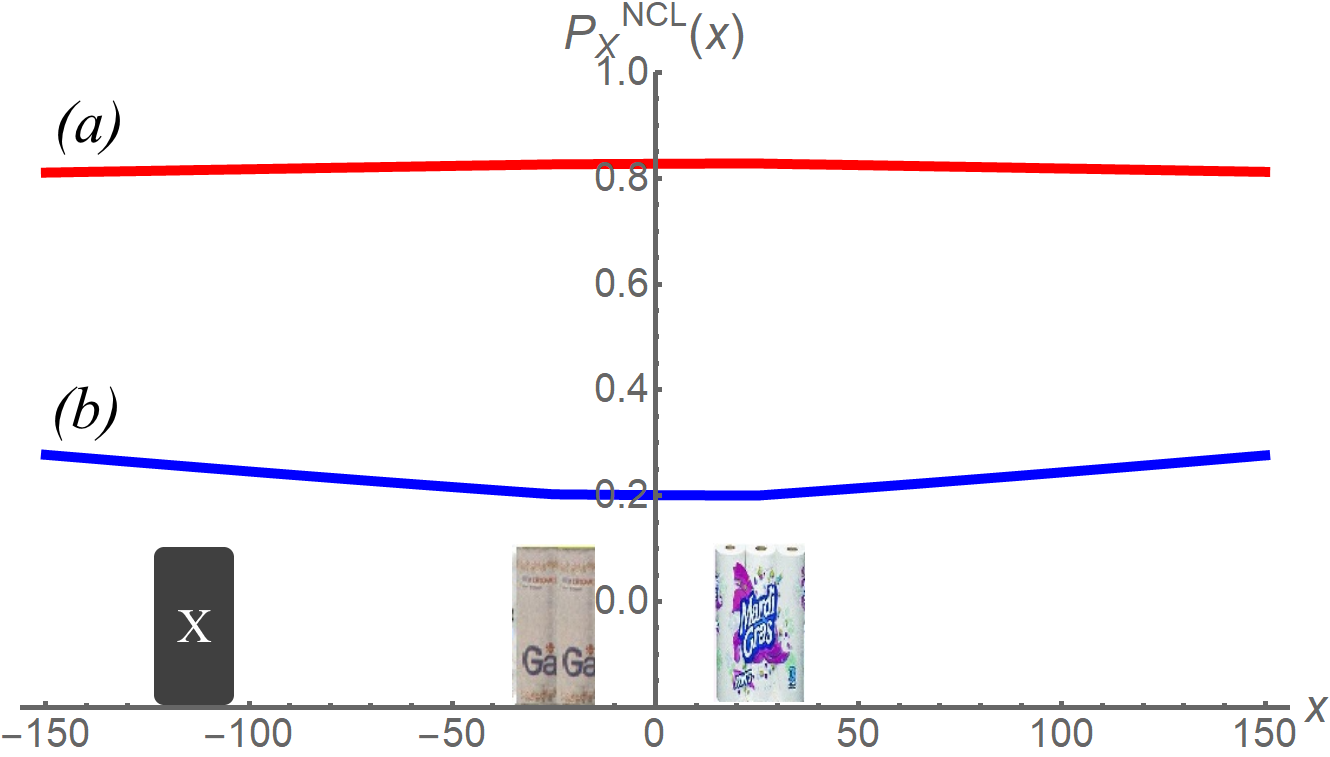}
        \caption{\footnotesize{Purchase probability using parameter estimates}}
        \label{diagram3prods}
    \end{subfigure}
    \begin{subfigure}[t]{0.47\textwidth} 
        \centering
        \includegraphics[width=0.95\textwidth]{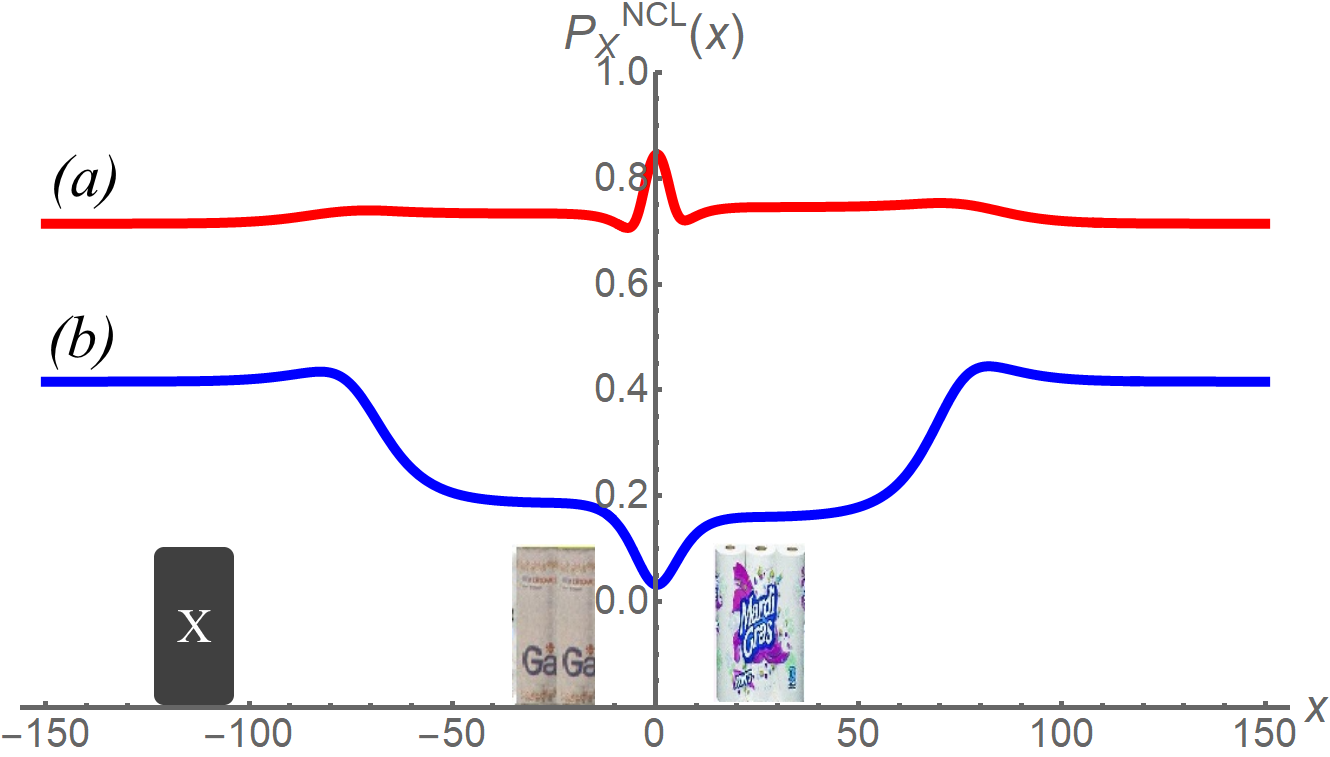}
        \caption{\footnotesize{Purchase probability using large $\gamma$}}
        \label{diagram3prodshighgamma}
    \end{subfigure}
    \caption{\footnotesize{Purchase probability of generic product $x$ as a function of its position on a shelf containing two other products GALA and MARDI GRA. On the left (a), using the estimated parameters (Table \ref{tab:PTW_BLP}), on the right (b) using a larger $\gamma$, which is equivalent to re-scaling shelf length.}} 
    \label{pp_three_prods}
\end{figure}

When all the products have low attractiveness, see case (b) of Figure \ref{diagram3prods}, we can see that if product $X$ is initially at the edge of the shelf ($x \ll -25$ or $x \gg +25$) and as it approaches the nearest product, the choice probability for product $X$ decreases slightly then flattens as it gets closer to the center between the competing products. This naturally highlights the increased competition that occurs when products of similar utility move closer to each other. The minimum occurs when product $X$ is sandwiched between the other two products leading to intense competition between all three products rather than only facing strong competition from one product when $X$ is on the end of the shelf. Interestingly, the effect is reverse when product $X$ is of high utility. As the product moves closer to the center and in between the two other products, $X$'s purchase probability increases. In this case, the negative effect of an increase in competition, is countered by the increase in visibility this product enjoys by being placed in the center.

In order to see the impact that $\gamma$ plays in generating these results, we run a comparative static that increases the value of $\gamma$ . Note, the increase in $\gamma$ is similar to shrinking the store shelf by the same factor and as a result the effects on purchase probability are more pronounced and nuanced. We present these results in Figure \ref{diagram3prodshighgamma}. For option b when all products are of low utility, being closer to a group makes product $X$ more noticeable to consumers which causes the initial increase in its demand. Notice that this effect eventually is overshadowed by the stronger negative competition effect from both products, which drives the demand for product $X$ down as it gets closer to the center. In fact, it is clear that the demand for product $X$ is lowest when product $X$ is exactly mid-way between the competing products. The reason is that product $X$ faces roughly equal competition from both products.  For case (a), similar but inverted and muted effects occur. 

When our target product is the dominant brand in the category, we observe that the choice probability $P^{NCL}_X$ peaks when the target product is in the center between the two competing products. At this point the dominant product is in a highly visible position relative to the competitors and effectively draws share away. We can see that as product $X$ moves away from the grouping of products towards the edge of the display that demand for $X$ begins to decrease. This drop in demand is most notable at the edges of the display, but there is a drop-off from being in the center and drawing share aware from both neighbors, and another drop off as it moves away from the nearest product.

\subsection*{Elasticity of NCL Model} \label{model_elas}

The previous section focused on the effect of position on the  choice probability as a function of the dissimilarity parameter and illustrates how utility is influenced by position. Typically, marketers also change the utility of products through price. In this section we consider the own- and cross-price elasticities as a function of the dissimilarity parameter. The importance of this analysis is that it shows that increased price substitution can be induced through shelf position.

\textbf{Elasticity with respect to product characteristics:} The own-characteristic elasticities $E^M_{j, \pmb{x}_{j}^\intercal\pmb{\beta}^c}$ and cross-characteristic elasticities $E^M_{j', \pmb{x}_{j}^\intercal\pmb{\beta}^c}$ of the model $M \in \{MNL, NL, HNCL\}$ are given in Table \ref{table:1}. The elasticities for MNL and NL models are reported for comparison. Observe that when the dissimilarity pattern reaches its maximum, $\rho=1$, that the HNCL's elasticities equals those of the MNL model. Otherwise both the own and cross-characteristic elasticities are higher  due to an extra factor that gets bigger when $\rho$ gets smaller. This increase reflects that nest-pairs are less similar and product competition outside the nest diminishes with respect to $j$. 


\begin{table}[H]
\caption{Elasticities associated with product characteristics for the MNL, NL, and NCL models.}
    \label{table:1}
      \centering
\resizebox{\textwidth}{!}{\begin{tabular}{r|cc}
        $M$ & Own-elasticity & Cross-elasticity\\
        \toprule
        MNL & $\displaystyle(1 - P^{MNL}_j)\pmb{x}^\intercal_j\pmb{\beta}^c$ & $\displaystyle -P^{MNL}_j\pmb{x}^\intercal_j\pmb{\beta}^c$\\
        NL & $\displaystyle\left(\left(\frac{1}{\rho} - 1\right)(1 - P_{j|\lbrace \sigma_j \rbrace}) + (1 - P^{NL}_j)\right)\pmb{x}^\intercal_j\pmb{\beta}^c$ & $\displaystyle-\left(\left(\frac{1}{\rho} - 1\right)P_{j|\lbrace \sigma_j \rbrace} + P^{NL}_j\right)\pmb{x}^\intercal_j\pmb{\beta}^c$\\
        HNCL & $\displaystyle\sum_{j'\neq j}\frac{P_{j|jj'}P_{jj'}}{P^{HNCL}_j}\left(\left(\frac{1}{\rho} - 1\right)(1 - P_{j|jj'}) + (1 - P^{HNCL}_j)\right)\pmb{x}^\intercal_j\pmb{\beta}^c$ & $\displaystyle-\left(\left(\frac{1}{\rho} - 1\right)\frac{P_{j|jj'}P_{jj'}P_{j'|jj'}}{P^{HNCL}_{j'}} + P^{HNCL}_j\right)\pmb{x}^\intercal_j\pmb{\beta}^c$\\
        \bottomrule
    \end{tabular}}
\end{table}

The own- and cross-characteristic elasticities for the NCL model can be written in term of the own- and cross-characteristic elasticities of HNCL model as follows:
\begin{align*}
    E^{NCL}_{k, \pmb{x}_{j}^\intercal\pmb{\beta}^c} = \frac{\pmb{x}_j^\intercal\pmb{\beta}^c}{P^{NCL}_{st}(k)}\int_{D_{st}\in \mathbb{R}}\int_{\pmb{v}_i \in \mathbb{R}^{J+1}}\frac{P^{NCL}_{ist}(k)}{\pmb{x}^\intercal_j\pmb{\beta}^c}E^{HNCL}_{k, \pmb{x}_j^\intercal\pmb{\beta}^c_{is}}dF(\pmb{v}_i)dF(D_{st})
p\end{align*}
where $k = j$ and $k = j'$ for own- and cross-characteristic elasticities respectively.

\textbf{Elasticity with respect to changes of the shelf position: } Traditional models in marketing are insensitive to changes in the product arrangements in the shelf. For example, when the effects are fixed they affect the levels of the probability but elasticities are not modified as the position changes. In the NCL model, a change in the shelf design can have multiple consequences to product demand. To simplify these interactions and to lead to a better understanding of the model's workings we consider how the choice probability of product $j$ varies when we change the location of a competing product $j'$ assuming everything else remains constant, i.e. every other distance between products in the shelf remain unchanged.  Furthermore, we look at the effects under the HNCL model to simply our analysis.

\begin{align*}\label{ela_shelf1}
 E^{HNCL}_{P_j,f_{jj'}}=
    &\frac{P_{j|jj'}P_{jj'}}{P^{HNCL}_j}\left(\frac{1 - \alpha_{j,jj'}}{\rho} + \left(\frac{1}{\rho} - 1\right)\left(\alpha_{j,jj'}P_{j|jj'} + \alpha_{j',jj'}P_{j'|jj'} - 1\right) - P_{jj'} + \alpha_{j,jj'}P^{HNCL}_j + \alpha_{j',jj'}P^{HNCL}_{j'}\right)\\
    &\qquad + \sum_{k\neq j,j'}\frac{P_{j|jk}P_{jk}}{P^{HNCL}_j}\left(-\frac{\alpha_{j,jj'}}{\rho} + \left(\frac{1}{\rho}-1\right)\alpha_{j,jj'}P_{j|jk} - P_{jj'} + \alpha_{j,jj'}P^{HNCL}_j + \alpha_{j',jj'}P^{HNCL}_{j'}\right)\numberthis
\end{align*}

This elasticity can be decomposed in two effects, when $j'$ is moving away from $j$, the competition between them decreases. For the other case, which is reflected in the first term, as moving away/closer means $\alpha_{j,jj'}$ and $\alpha_{j',jj'}$ get smaller/bigger. The second effect is that competition between $j$ and other products in the shelf increase/decrease when $j'$ moves away/closer, as now the perceived distance from $j$ to the other products in the shelf is lower/higher.

\begin{figure}[ht]
    \centering
    \includegraphics[width=0.49\textwidth]{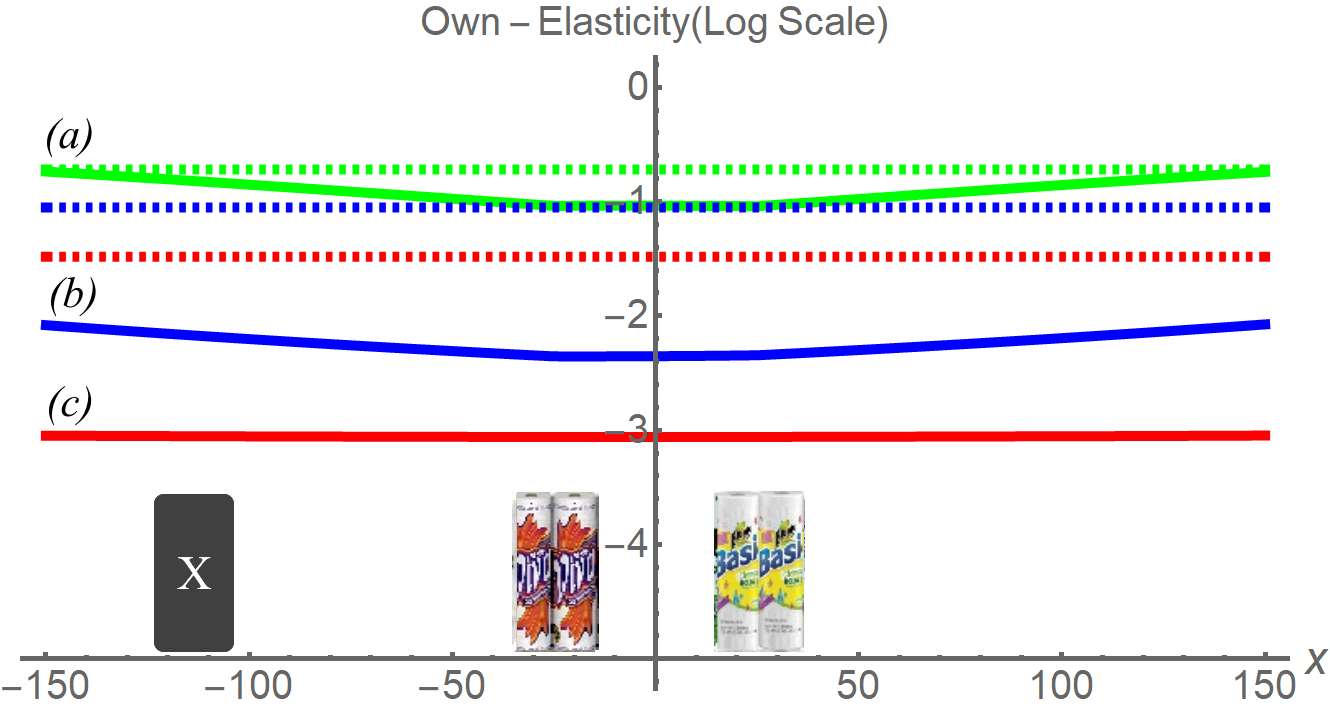}
    \includegraphics[width=0.49\textwidth]{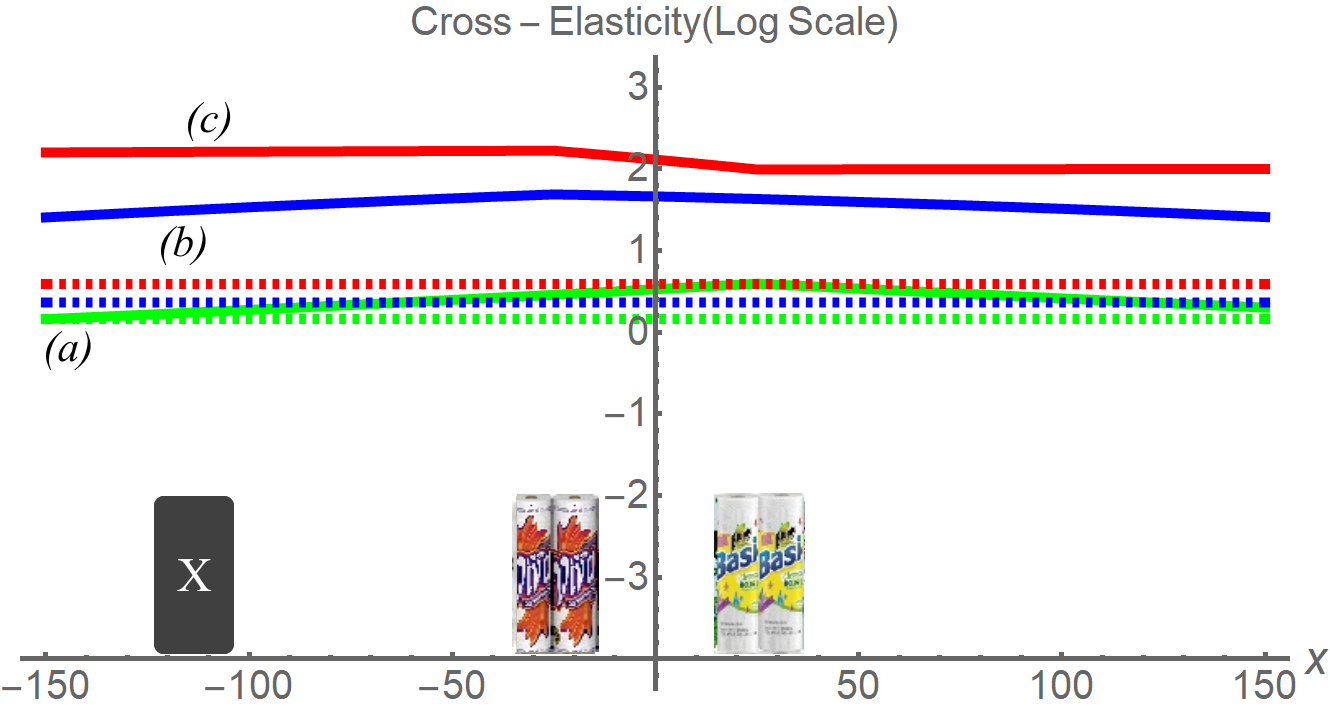}
    \caption{\footnotesize{Linear shelf that contains two fixed products VIVA and BOUNTY. Green, blue and red represent Own and Cross price--elasticities for products (a) DOMINICK'S, (b) CORONET and (c) BRAWNY, with high, medium and low utility products respectively. Solid line is computed using NCL model, while dotted line is computed using MNL model.}}
    \label{diagram3prodselasticity}
\end{figure}

\textbf{Example} To illustrate the effect of shelf position on elasticities, let us consider a setting similar to the one presented in the illustrative example in figure \ref{pp_three_prods}. In this setting the shelf is represented as a single row with three products. We fix two product with average value to consumers, VIVA and DOMINICK'S located at -25 and +25 inches from the center of the single row shelf.  Additionally, we introduce a third product with either a high, medium, or low product value to customers (DOMINICK'S, CORONET and BRAWNY marked in red, blue and green respectively) and we call this latter product $X$. Here we analyze how the own-price-elasticity of product $X$ varies depending on its location on the shelf as shown in Figure \ref{diagram3prodselasticity}. On the left, we observe that the own price elasticity of $X$ is closer to zero, more inelastic, when it is far from other products and it is at its lowest, more elastic, when it is closest to both fixed products, in the center of the shelf, signaling that competition effect dominates on the three cases. Interestingly, we also observe that the effect on the price elasticities is stronger in the case of the medium value product. When the product utility of product $x$ is too low or too high compared to the fixed products, the difference in price elasticity is (a) 0.3\% (c) 1.3\% respectively, but when the value is comparable, the change in price elasticity is (b) 32.3\%, in our example CORONET would really benefit from choosing a location far from VIVA and BOUNTY, while DOMINICK'S and BRAWNY would not be affected much by changing locations in this linear shelf. We also included the MNL implied elasticities in dotted lines, which are not affected by the position of the product in the shelf.


 A similar result is observed when we look at the cross-price elasticity on the right, where we observe the positive effect on the cross-price elasticiy when moving $x$ on the shelf. We can see from Figure \ref{diagram3prodselasticity} that the effect is no longer symmetric because the cross--price elasticity is with respect to the price of VIVA (the product on -25). In all three cases of the utility associated with product $X$, we see the cross--price elasticity increases as it approaches Viva. For the case of the high and medium utility product, as product $X$ passes VIVA, the cross--price elasticity declines. However, for the  case of the low utility products $X$ the cross price utility increases highlighting the importance of shelf position on own and cross--price elasticities. The absolute effect on DOMINICK's with higher utility, is small, but for the medium and low utility products CORONET and BRAWNY the effect is larger and Again the most affected product from this movement is CORONET, with a utility similar to the fixed products VIVA and BOUNTY. The ratio between the lowest and highest cross--price elasticities for CORONET is 31.5\% and for BRAWNY is 25.8\%, which shows a significant variation.

\begin{figure}[ht]
    \centering
    \includegraphics[width=0.49\textwidth]{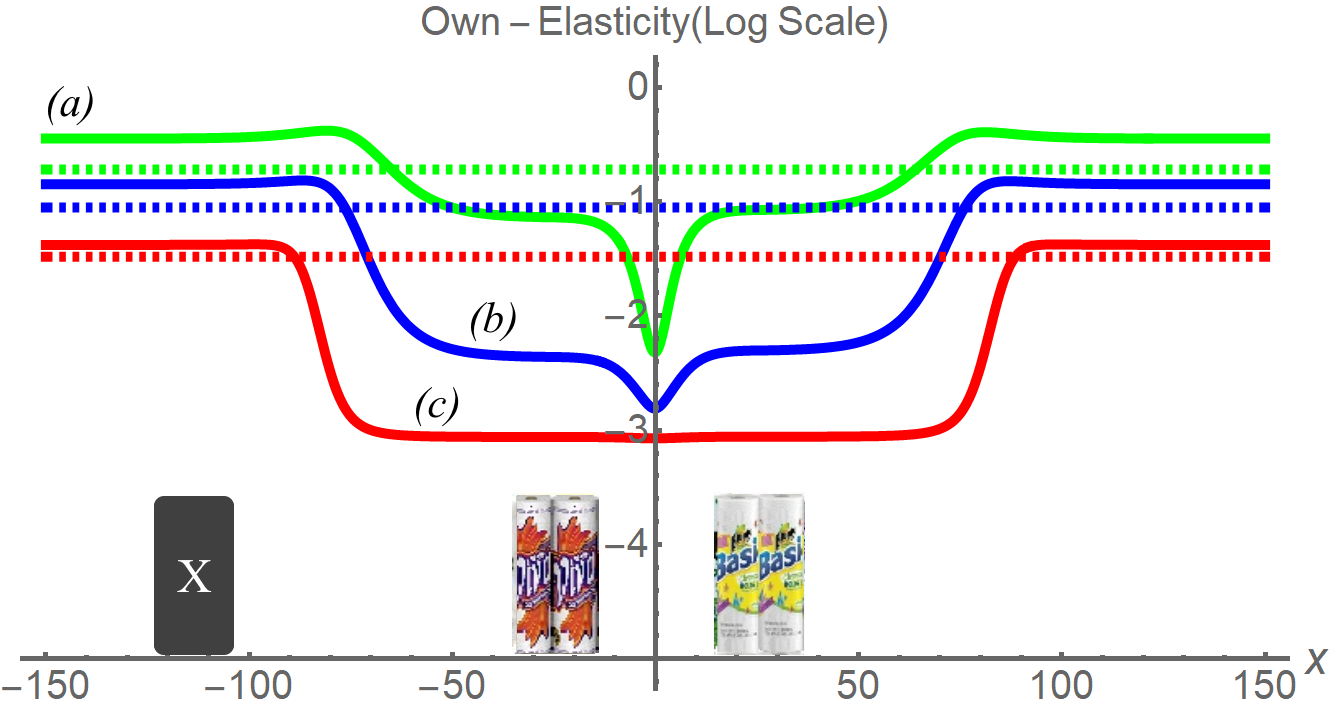}
    \includegraphics[width=0.49\textwidth]{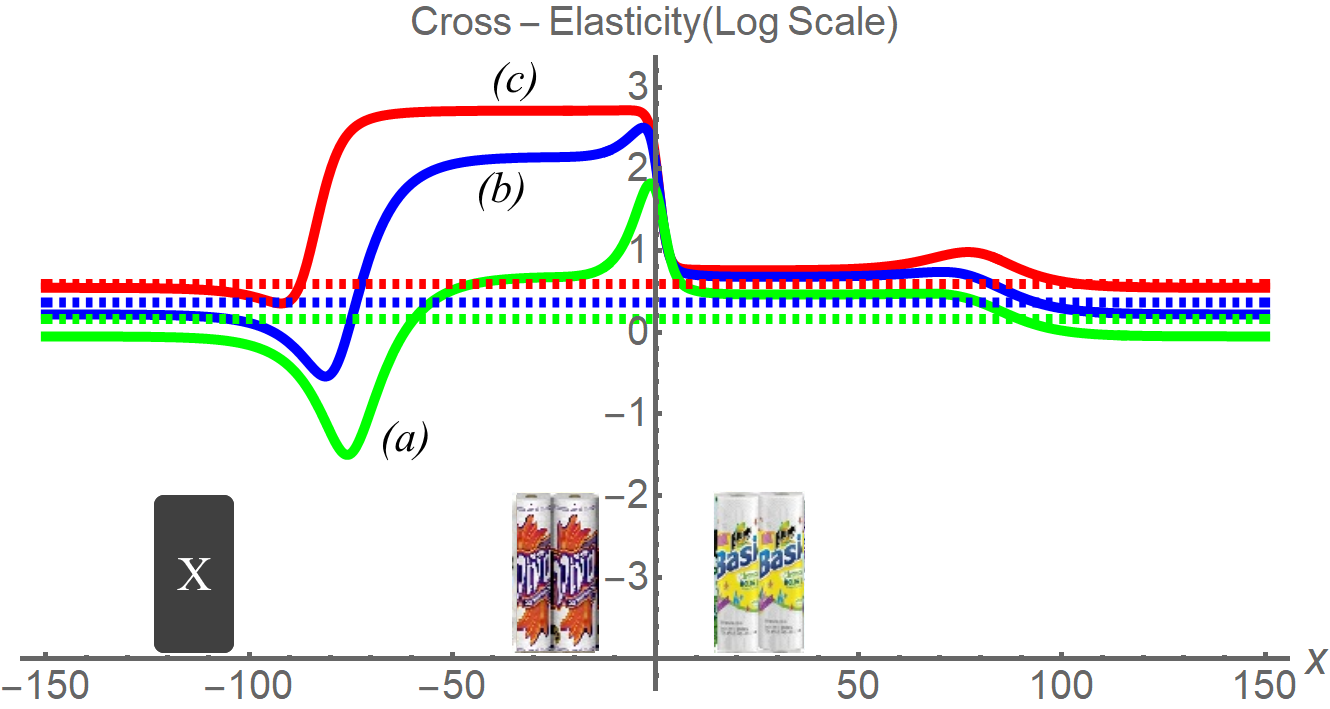}
    \caption{\footnotesize{We repeat the same plot as Figure \ref{diagram3prodselasticity} increasing the value of $\gamma$. This is equivalent to rescaling the shelf in Figure \ref{diagram3prods} horizontally. Solid line are high, medium and low utility products (a,b and c) and the dotted lines are the corresponding MNL implied elasticities.}}
    \label{diagram3prodselasticity50times}
\end{figure}

In order to glean more insight from our example, we run a comparative statics with respect to gamma, increasing its value, which is presented in Figure \ref{diagram3prodselasticity50times}. It can be seen from the graph that the own--elasticity is higher in the most price elastic position compared to the least. Similarly, the cross--elasticity is higher in the most price elastic position compared to the least. Note when the utility of $X$ is high, the increase in the price of VIVA has a stronger positive effect on the demand of $X$ on the left side of the origin compared to on the right as the competitiveness of the VIVA located in the close proximity diminishes. On the other hand, if the utility of $X$ is low, the positive effect on the demand of $X$ is weaker on the left side of the origin compared to on the right. This observation is consistent with the fact that $X$ with low utility is a competitive product, hence its demand benefits from the visibility effect by being close to another attractive product VIVA. Since the increasing price of VIVA affects is negative on the visibility of product $X$, the overall demand of $X$ responds less positively on the left of the origin compared to on the right.

We can also compare the own-- and cross--price elasticities of $X$ according to the NCL model to that of the MNL model (shows in dashed line in Figure \ref{diagram3prodselasticity} and Figure \ref{diagram3prodselasticity50times}). The price--elasticities are higher for NCL model compared to the MNL when $X$ is closer to other products. This can be understood as the dissimilarity parameter $\rho < 1$ for the NCL model indicates that products are quite good substitute of each other. However when we increase the magnitude of $\gamma$ by $50$, we can see that when $X$ is located far away from product $L$ and $R$, it is less price--elastic according to the NCL model compared to the MNL model, as $X$ is getting compared less to other products. Another interesting feature revealed is when $X$ is a medium of high utility product (as in the (a) and (b) cases in Figure \ref{diagram3prodselasticity50times}), there is a finite position where the elasticities are minimized. This is the optimal points where $X$ is most visible but does not suffer too high competition, hence it is less affect by price fluctuations.  

Another interesting property of the NCL model is when the change in the shelf happens in a remote location, there are ripple effects in the demand of the unchanged products due to variation in competition levels induced by the shelf. This is depicted in equation (\ref{ela_shelf2}) 

\begin{equation}\label{ela_shelf2}
        E^{HNCL}_{P_k,f_{jj'}}=
        \left(\frac{1}{\rho} - 1\right)\frac{\alpha_{j,jj'}P_{j|jk}P_{jk}P_{k|jk}}{P^{HNCL}_k} + \left(\frac{1}{\rho} - 1\right)\frac{\alpha_{j',jj'}P_{j'|j'k}P_{j'k}P_{k|j'k}}{P^{HNCL}_k} - P_{jj'} + \alpha_{j,jj'}P^{HNCL}_j + \alpha_{j',jj'}P^{HNCL}_{j'}.
\end{equation}

The corresponding elasticities for the NCL model can be obtained from the elasticities of HNCL model as follows:
\begin{align*}
    E^{NCL}_{k, f_{jj'}} = \frac{1}{P^{NCL}_{st}(k)}\int_{D_{st}\in \mathbb{R}}\int_{\pmb{v}_i \in \mathbb{R}^{J+1}}P^{NCL}_{ist}(k)E^{HNCL}_{k, f_{jj'}}dF(\pmb{v}_i)dF(D_{st}).
\end{align*}
Where we may set $k = j$ to obtain the heterogeneous counter-part of (\ref{ela_shelf1}).




\section*{Increased Profits through Improved Shelf Design} \label{designandprofitsection} 

Our research shows that a key aspect of shelf design is the possibility that location influences the substitution rate between products. Therefore the decision of where to place each product on the shelf in order to maximize profits depends upon: (1) the product's intrinsic value driven by its design, (2) extrinsic value drivers such as price, (3) the absolute quality of the shelf location, as well as (4) the relative value of the shelf location based upon the position of all the other products on the shelf. In previous sections we discuss the relative shelf position effects on demand, albeit in a stylized manner. In this section we turn our attention to the impact shelf design has on profit by employing our estimated model and the sales data from Dominick's Finer Foods.    

We begin by evaluating the four shelf designs that were implemented in the micro-marketing study. The first design organized products by quality and size by placing premium single rolls on the top shelf (providing high visibility), multi-counts in the middle, and low price brands on the bottom shelf. The second design organized shelves by placing low and mid price single rolls on the top shelf, multi-packs in the middle, and premium on the bottom. The third design made price comparisons very difficult by vertically merchandising single rolls and multi-packs. The last design and the study's control group merchandised all sizes within brand blocks. Again, these designs are illustrated in Figure \ref{fig:planograms}. 

Below in Figure \ref{fig:profits} we present the expected profit per purchase associated with designs 1 to 4. The data give us product margins by week and store. We thus determine the expected profit margin by multiplying the product margin times the probability of purchase average over weeks and stores. As illustrated, designs 1 and 2 are determined to be more profitable than Design 3 and the control. The intuition for such a result lies with the designs of 1 and 2, which leverage the benefit of placement of other nearby high value products. Design 3 is the worst shelf design, because the relative shelf competition is reduced by making price comparisons very difficult across quality and size and the visibility effect is reduced from product placement on the shelf.

\begin{table}[ht]
    \centering
\begin{tabular}{c|ll}
    
        Model & $NCL^{exp}$ \\
        \toprule
        Control & 0.198009 \\
        Design 1 & 0.198711 (+0.35\%) \\
        Design 2 & 0.199283 (+0.64\%)\\
        Design 3 & 0.195310 (-1.36\%)\\
        New Design A &  0.204635 (+3.35\%)\\
        Best random design &  0.204895 (+3.48\%)\\
        Worst random design &  0.191115 (-3.48\%)\\
        Average random design &   0.197474 (-0.27\%)\\
        \bottomrule
    \end{tabular}
\caption{\footnotesize{The table shows the expected dollar amount of profit (\$) per purchase of each design in the PTW category. The plot (a) shows the expected dollar amount of profit (\$) per item of 100,000 randomly generated shelf designs according to the $NCL^{exp}$ model.}}
\label{fig:profits}
\end{table}

\begin{figure}[ht]
    \centering
    \includegraphics[scale=0.6]{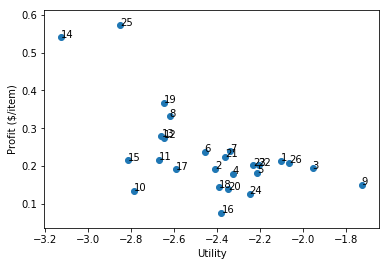}
    \caption{\footnotesize{Plot of utility vs. profit for the 26 products in the Paper Towel Category.}}
    \label{fig:uvsprofit}
\end{figure}

We expand our discussion to an alternative store shelf design using Figure \ref{fig:uvsprofit} which suggests a negative correlation between profitability and utility in the paper towel category. This shows that the paper towel category can be roughly considered as a set of high utility and low profit products with low utility, high profit products. For example, in Figure \ref{fig:uvsprofit}, product 10 and product 25 have similar utility, but product 25 is more profitable, while product 1 has a much higher utility and profitability than product 16's.
Using Figure \ref{fig:uvsprofit} in conjunction with the attractiveness score map of Figure \ref{fig:shelfwithattr}, we develop an alternative design that increases the expected profit, according to $NCL^{exp}$ model. This alternative design puts high profit low utility products, such as products 14 and 25, in the center of the shelf where the attractiveness is high. High utility and (relatively) high profit products, such as 1, 26, and 3, are placed at the ends of the shelf to maximize the distance between each category. The rest of the shelf is filled with low utility low profit products, such as products 10 and 15, as a buffer between the center and ends of the shelf.

We refer to our new design for the $NCL^{exp}$ model as New Design A. Figure \ref{fig:profits} shows the expected profit per purchase associated with designs 1 to 3, the control, and our New Design A. The $NCL^{exp}$ model predicts an improvement of +3.35\% using New Design A compared to our control design. These expected profit increases are on gross profit and quite substantial, given that there is only a one-time cost of rearranging items on the shelf. In order to put this new design into perspective, we compare it to 100,000 randomly generated shelf designs based on 100,000 random permutations of products from the control design. Figure \ref{fig:profits} shows how the expected profits per purchase of randomly generated designs are distributed. We can see that the expected dollar amount of profit per purchase of our New Design A is about $3.8\sigma$ higher than in the average random design, according to the $NCL^{exp}$ model. We further note that the profit per purchase of our New Design A is +7.07\% larger than the respective worst random design, according to $NCL^{exp}$, which emphasizes the importance of choosing a better shelf design.

Our approach to shelf design for profit improvement as above has been heuristic, and there is no guarantee we will arrive at the optimal design. Even for our relatively small example shelf with 26 products, we can see that it is already a complex problem as there are 26! possible designs to consider. This highlights one important difficulty with shelf design optimization: the complexity tends to increase rapidly with the number of products. One key strength of our model is its capability to capture the essence of both absolute and relative position effects with relatively simple closed--form choice probabilities. On the other extreme, it is also possible to consider a choice probability model where a consumer first decide among $2^{26}$ possible consideration sets, but the model would be complicated and most importantly the time--complexity of choice probability evaluation would be exponential in the number of products. The simplicity of our model becomes important when formulating our problem to be solved using standard non--linear integer programming tools or software, where fast choice probability computation can helps speed--up the overall run--time of the optimization algorithm.

\section*{Application to a Virtual Store}\label{sec:virtual_store}

 It is interesting to see if the findings can be generalized to the digital world. We conduct a lab experiment to collect choice data from digital shelves. In this setting our ``shelf'' is a computer screen in which the products are organized as either a list or a grid.  We create an experiment in which participants look at different product arrangement and make choices. The number of products ranges from 9 to 12.  We tested the model for 6 different categories, with different degrees of brand loyalty, according to a pretest study.

We hypothesize that the same effect can take place in the digital world. In a screen display, products still need to be shown to consumers in a specific order, and in some cases, such as at Amazon or Walmart, the number of products can be much larger than in a physical store, making this effect potentially even more prominent.

\begin{table}[ht]
    \centering
\resizebox{0.8\textwidth}{!}{\begin{tabular}{lcccccccc}
     \toprule
 & $NCL^{exp\dagger}$ & $MNL^\dagger$ & $NCL^{inv\dagger}$ & $NCL^{adj\dagger}$ & $NCL^{inv}$ & $MNL$ & $NCL^{adj}$ & $NCL^{exp}$\\ 
 \hline 
CB list & \color[rgb]{0.2857142857142857,0.7142857142857143,0} 8.58\% & \color[rgb]{0.0,1.0,0} 0.00\% & \color[rgb]{0.14285714285714285,0.8571428571428572,0} 8.36\% & \color[rgb]{0.42857142857142855,0.5714285714285714,0} 9.89\% & \color[rgb]{0.7142857142857143,0.2857142857142857,0} 41.04\% & \color[rgb]{0.8571428571428571,0.1428571428571429,0} 48.48\% & \color[rgb]{0.5714285714285714,0.4285714285714286,0} 36.61\% & \color[rgb]{1.0,0.0,0} 51.22\%\\ 
CB grid & \color[rgb]{0.0,1.0,0} 0.00\% & \color[rgb]{0.42857142857142855,0.5714285714285714,0} 4.38\% & \color[rgb]{0.2857142857142857,0.7142857142857143,0} 3.17\% & \color[rgb]{0.14285714285714285,0.8571428571428572,0} 1.80\% & \color[rgb]{0.5714285714285714,0.4285714285714286,0} 57.37\% & \color[rgb]{0.8571428571428571,0.1428571428571429,0} 68.02\% & \color[rgb]{0.7142857142857143,0.2857142857142857,0} 60.87\% & \color[rgb]{1.0,0.0,0} 70.30\%\\ 
 \hline 
SD list & \color[rgb]{0.2857142857142857,0.7142857142857143,0} 4.10\% & \color[rgb]{0.0,1.0,0} 0.00\% & \color[rgb]{0.14285714285714285,0.8571428571428572,0} 3.17\% & \color[rgb]{0.42857142857142855,0.5714285714285714,0} 4.37\% & \color[rgb]{0.8571428571428571,0.1428571428571429,0} 42.01\% & \color[rgb]{0.5714285714285714,0.4285714285714286,0} 38.13\% & \color[rgb]{1.0,0.0,0} 45.94\% & \color[rgb]{0.7142857142857143,0.2857142857142857,0} 41.30\%\\ 
SD grid & \color[rgb]{0.14285714285714285,0.8571428571428572,0} 5.94\% & \color[rgb]{0.0,1.0,0} 0.00\% & \color[rgb]{0.2857142857142857,0.7142857142857143,0} 6.24\% & \color[rgb]{0.42857142857142855,0.5714285714285714,0} 7.01\% & \color[rgb]{0.8571428571428571,0.1428571428571429,0} 80.43\% & \color[rgb]{1.0,0.0,0} 88.30\% & \color[rgb]{0.5714285714285714,0.4285714285714286,0} 63.32\% & \color[rgb]{0.7142857142857143,0.2857142857142857,0} 76.98\%\\ 
 \hline 
CH list & \color[rgb]{0.0,1.0,0} 0.00\% & \color[rgb]{0.2857142857142857,0.7142857142857143,0} 0.78\% & \color[rgb]{0.42857142857142855,0.5714285714285714,0} 3.59\% & \color[rgb]{0.14285714285714285,0.8571428571428572,0} 0.53\% & \color[rgb]{0.8571428571428571,0.1428571428571429,0} 28.01\% & \color[rgb]{1.0,0.0,0} 31.04\% & \color[rgb]{0.7142857142857143,0.2857142857142857,0} 27.71\% & \color[rgb]{0.5714285714285714,0.4285714285714286,0} 26.04\%\\ 
CH grid & \color[rgb]{0.2857142857142857,0.7142857142857143,0} 2.54\% & \color[rgb]{0.14285714285714285,0.8571428571428572,0} 1.16\% & \color[rgb]{0.0,1.0,0} 0.00\% & \color[rgb]{0.42857142857142855,0.5714285714285714,0} 3.39\% & \color[rgb]{0.8571428571428571,0.1428571428571429,0} 27.99\% & \color[rgb]{0.5714285714285714,0.4285714285714286,0} 27.44\% & \color[rgb]{0.7142857142857143,0.2857142857142857,0} 27.76\% & \color[rgb]{1.0,0.0,0} 29.51\%\\ 
 \hline 
CK list & \color[rgb]{0.0,1.0,0} 0.00\% & \color[rgb]{0.2857142857142857,0.7142857142857143,0} 3.80\% & \color[rgb]{0.14285714285714285,0.8571428571428572,0} 0.79\% & \color[rgb]{0.42857142857142855,0.5714285714285714,0} 4.53\% & \color[rgb]{0.5714285714285714,0.4285714285714286,0} 46.15\% & \color[rgb]{0.8571428571428571,0.1428571428571429,0} 49.10\% & \color[rgb]{1.0,0.0,0} 50.39\% & \color[rgb]{0.7142857142857143,0.2857142857142857,0} 46.18\%\\ 
CK grid & \color[rgb]{0.14285714285714285,0.8571428571428572,0} 0.00\% & \color[rgb]{0.42857142857142855,0.5714285714285714,0} 0.00\% & \color[rgb]{0.2857142857142857,0.7142857142857143,0} 0.00\% & \color[rgb]{0.0,1.0,0} 0.00\% & \color[rgb]{1.0,0.0,0} 14.94\% & \color[rgb]{0.5714285714285714,0.4285714285714286,0} 1.30\% & \color[rgb]{0.7142857142857143,0.2857142857142857,0} 3.58\% & \color[rgb]{0.8571428571428571,0.1428571428571429,0} 3.85\%\\ 
 \hline 
CL list & \color[rgb]{0.42857142857142855,0.5714285714285714,0} 1.71\% & \color[rgb]{0.14285714285714285,0.8571428571428572,0} 0.00\% & \color[rgb]{0.2857142857142857,0.7142857142857143,0} 0.78\% & \color[rgb]{0.0,1.0,0} 0.00\% & \color[rgb]{0.5714285714285714,0.4285714285714286,0} 34.56\% & \color[rgb]{0.8571428571428571,0.1428571428571429,0} 40.60\% & \color[rgb]{0.7142857142857143,0.2857142857142857,0} 39.40\% & \color[rgb]{1.0,0.0,0} 42.61\%\\ 
CL grid & \color[rgb]{0.2857142857142857,0.7142857142857143,0} 1.11\% & \color[rgb]{0.0,1.0,0} 0.00\% & \color[rgb]{0.42857142857142855,0.5714285714285714,0} 1.45\% & \color[rgb]{0.14285714285714285,0.8571428571428572,0} 0.97\% & \color[rgb]{0.5714285714285714,0.4285714285714286,0} 71.90\% & \color[rgb]{0.7142857142857143,0.2857142857142857,0} 72.53\% & \color[rgb]{1.0,0.0,0} 75.22\% & \color[rgb]{0.8571428571428571,0.1428571428571429,0} 72.60\%\\ 
 \hline 
DT list & \color[rgb]{0.14285714285714285,0.8571428571428572,0} 0.11\% & \color[rgb]{0.0,1.0,0} 0.00\% & \color[rgb]{0.42857142857142855,0.5714285714285714,0} 1.72\% & \color[rgb]{0.2857142857142857,0.7142857142857143,0} 0.27\% & \color[rgb]{0.7142857142857143,0.2857142857142857,0} 46.20\% & \color[rgb]{0.8571428571428571,0.1428571428571429,0} 53.56\% & \color[rgb]{0.5714285714285714,0.4285714285714286,0} 36.77\% & \color[rgb]{1.0,0.0,0} 56.80\%\\ 
DT grid & \color[rgb]{0.0,1.0,0} 0.00\% & \color[rgb]{0.14285714285714285,0.8571428571428572,0} 0.53\% & \color[rgb]{0.2857142857142857,0.7142857142857143,0} 0.87\% & \color[rgb]{0.42857142857142855,0.5714285714285714,0} 2.99\% & \color[rgb]{0.5714285714285714,0.4285714285714286,0} 79.48\% & \color[rgb]{0.8571428571428571,0.1428571428571429,0} 86.90\% & \color[rgb]{1.0,0.0,0} 89.82\% & \color[rgb]{0.7142857142857143,0.2857142857142857,0} 81.77\%\\ 
    \bottomrule
    \end{tabular}}
    \caption{\footnotesize{RMSE relative to the best model in testing data}}
    \label{tab:Screen_test}
\end{table}


We used the two most common types of product arrangements used in online retailing: a list of products (either in a column or a row) and a matrix with 9 to 12 products shown simultaneously. We did not include displays that spanned for multiple screens, so participants did not need to scroll down, to avoid other considerations in the experiment. This setting challenges the findings, because the number of products is small, and consumer do not need to walk or scroll to other pages. We hypothesize that the effect should be even stronger when consumers can see only a subset of products at a time or the assortment includes more products. The model predictive performance can be seen in Table \ref{tab:Screen_test}.

We tested the categories Candy Bar (CB), Soft Drinks (SD), Cheese (CH), Cookies (CK), Detergent (DT), and an additional experiment with colors, to avoid brand loyalty and other effects. Here we can observe that the NCL model has a good performance, even in the low product number and all products in a single screen. This result supports our argument that relative shelf position affects consumer choices. When products are organized in a list, we can still find shelf effects, but the effect looks stronger when products are organized in matrix form.

\section*{Conclusion and Managerial Implications}\label{sec:conclusions}

In this paper, we introduce the NCL model to capture the influence of shelf position on demand. Specifically, we introduce both absolute and relative shelf position effects. Absolute effects refer to good shelf locations such as eye-level positions that improve the attractiveness of any product in that position. Relative effects refer to the competitive effects of products nearby on the shelf and depend upon the relative attractiveness of the focal product versus the attractiveness of its neighbors.

We show that this NCL model fits better than traditional choice models. Moreover, it can be used to predict improvements in sales and profitability from improved shelf design. This model is parsimonious and offers meaningful parameters. When we fit the model to real purchases, the parameter $\rho$ is significantly different from 1. This suggests that the effect of the shelf on demand is strong and that without the NCL's spatial component of shelf competition, there is a significant bias in the price sensitivity. Biased price sensitivity can lead to poor decisions by retailers and manufacturers about pricing and shelving. In our data, we show that this price sensitivity parameter has a bias of around 20$\%$ for the paper towel category. The parameter $\gamma$ measures the importance of distance between products through the allocation parameter $\alpha$; it is also significant with the right sign, meaning that the separation plays a role in the product competition.

Relative effects may be positive or negative depending upon two effects: competition and visibility. When products are located near each other, intuitively, they are compared more often, thus intensifying the competition between them, which can have a negative effect on demand. Yet, being compared more often may increase a product's visibility, which has a positive effect on product demand. Depending upon the relative attractiveness of the product, the combination of these two effects may be positive or negative. In particular, when a product yields high utility to consumers, the visibility effect dominates the competition effect. This implies that products with high value for consumers should be placed in prominent locations in shelves, and neighbor other highly attractive products. Analogously, products with low utility will prefer being placed in a neighborhood with other low utility products. Interestingly, in our data, products with intermediate value for customers are the most sensitive to changes in location and their price elasticities can change up 30\% from the worse location to the best.

Consequently, spreading out the locations of high utility and highly profitable products throughout the shelf will likely yield higher expected profit. The expected profit can be enhanced further by the visibility effect if it is possible to surround the location of each high utility profitable product with other products that have above average utility but are not as profitable. Conversely, the expected profit can also be increased by placing low utility but highly profitable products with other low utility products. We evaluate these recommendations by creating a new shelf design and using our fitted model. This analysis shows that an improvement of $3.35\%$ can be achieved by choosing a different shelf design that considers competition and visibility effects.

This research offers insight into how products compete spatially and how these spatial effects in turn can moderate or enhance competition. We show that this spatial aspect is relevant and important to understand as a separate effect from a location's attractiveness and product assortment. The model is applicable beyond a physical context and could be used for web design or virtual store design. For instance, on a screen display, products still need to be shown to consumers in a specific order, and in some cases, such as at Amazon or Walmart, the number of products can be much larger than in a physical store, making the spatial aspect potentially even more prominent.

We point out limitations of our research. We do not observe the consideration sets nor individual level behavior; therefore, we cannot observe competitive and visibility effects directly. Moreover, our experiments pertain to one retailer, and we cannot guarantee that our empirical findings will generalize to other cases. The distance functions that we chose are not exhaustive, and perhaps alternative ways of differentiating vertical and horizontal spacing could be considered. In our application, the product assortments are the same across stores, but if assortments vary, then they require different shelf designs. Therefore, product shelf and assortment could be jointly optimized. Finally, the neighborhood of products represents a challenging question due to the intrinsic combinatorial aspect of permutation that arises from testing every configuration. Each of these limitations provides new directions for future investigation.

\bibliographystyle{abbrvnat}
\bibliography{references}

\begin{thebibliography}{44}
\providecommand{\natexlab}[1]{#1}
\providecommand{\url}[1]{\texttt{#1}}
\expandafter\ifx\csname urlstyle\endcsname\relax
  \providecommand{\doi}[1]{doi: #1}\else
  \providecommand{\doi}{doi: \begingroup \urlstyle{rm}\Url}\fi

\bibitem[Agarwal et~al.(2011)Agarwal, Hosanagar, and
  Smith]{agarwal2011location}
A.~Agarwal, K.~Hosanagar, and M.~D. Smith.
\newblock Location, location, location: An analysis of profitability of
  position in online advertising markets.
\newblock \emph{Journal of Marketing Research}, 48\penalty0 (6):\penalty0
  1057--1073, 2011.

\bibitem[Anderson(1979)]{anderson1979analysis}
E.~E. Anderson.
\newblock An analysis of retail display space: theory and methods.
\newblock \emph{Journal of Business}, pages 103--118, 1979.

\bibitem[Bell and Song(2007)]{bellsong2007}
D.~R. Bell and S.~Song.
\newblock Neighborhood effects and trial on the internet: Evidence from online
  grocery retailing.
\newblock \emph{Quantitative Marketing and Economics}, 5:\penalty0 361--400,
  2007.

\bibitem[Ben-Akiva and Lerman(1977)]{ben1977disaggregate}
M.~Ben-Akiva and S.~Lerman.
\newblock Disaggregate travel demand and mobility choice models and measures of
  accessibility.
\newblock In \emph{3rd Internat. Conf. on Behavioral Demand Modeling}, 1977.

\bibitem[Berry et~al.(1995)Berry, Levinsohn, and Pakes]{berry1995automobile}
S.~Berry, J.~Levinsohn, and A.~Pakes.
\newblock Automobile prices in market equilibrium.
\newblock \emph{Econometrica: Journal of the Econometric Society}, pages
  841--890, 1995.

\bibitem[Bhat(1998)]{bhat1998analysis}
C.~R. Bhat.
\newblock Analysis of travel mode and departure time choice for urban shopping
  trips.
\newblock \emph{Transportation Research Part B: Methodological}, 32\penalty0
  (6):\penalty0 361--371, 1998.

\bibitem[Bhat and Guo(2004)]{bhat2004mixed}
C.~R. Bhat and J.~Guo.
\newblock A mixed spatially correlated logit model: formulation and application
  to residential choice modeling.
\newblock \emph{Transportation Research Part B: Methodological}, 38\penalty0
  (2):\penalty0 147--168, 2004.

\bibitem[Bianchi-Aguiar et~al.(2018)Bianchi-Aguiar, Silva, Guimar{\~a}es,
  Carravilla, and Oliveira]{bianchi2018allocating}
T.~Bianchi-Aguiar, E.~Silva, L.~Guimar{\~a}es, M.~A. Carravilla, and J.~F.
  Oliveira.
\newblock Allocating products on shelves under merchandising rules: Multi-level
  product families with display directions.
\newblock \emph{Omega}, 76:\penalty0 47--62, 2018.

\bibitem[Bradlow et~al.(2005)Bradlow, Bronnenberg, Russell, Arora, Bell,
  Duvvuri, ter Hofstede, Sismeiro, Thomadsen, and Yang]{bradlow2005}
E.~T. Bradlow, B.~Bronnenberg, G.~J. Russell, N.~Arora, D.~R. Bell, S.~D.
  Duvvuri, F.~ter Hofstede, C.~Sismeiro, R.~Thomadsen, and S.~Yang.
\newblock Spatial models in marketing.
\newblock \emph{Marketing Letters}, 16:3/4:\penalty0 267--278, 2005.

\bibitem[Bresnahan et~al.(1996)Bresnahan, Stern, and
  Trajtenberg]{bresnahan1996market}
T.~F. Bresnahan, S.~Stern, and M.~Trajtenberg.
\newblock Market segmentation and the sources of rents from innovation:
  Personal computers in the late 1980's.
\newblock Technical report, National Bureau of Economic Research, 1996.

\bibitem[Bucklin and Sismeiro(2003)]{bucklin2003model}
R.~E. Bucklin and C.~Sismeiro.
\newblock A model of web site browsing behavior estimated on clickstream data.
\newblock \emph{Journal of Marketing Research}, 40\penalty0 (3):\penalty0
  249--267, 2003.

\bibitem[Bultez and Naert(1988)]{bultez1988sh}
A.~Bultez and P.~Naert.
\newblock Sh. arp: Shelf allocation for retailers' profit.
\newblock \emph{Marketing Science}, 7\penalty0 (3):\penalty0 211--231, 1988.

\bibitem[Chandon et~al.(2009)Chandon, Hutchinson, Bradlow, and
  Young]{chandon2009does}
P.~Chandon, J.~W. Hutchinson, E.~T. Bradlow, and S.~H. Young.
\newblock Does in-store marketing work? effects of the number and position of
  shelf facings on brand attention and evaluation at the point of purchase.
\newblock \emph{Journal of Marketing}, 73\penalty0 (6):\penalty0 1--17, 2009.

\bibitem[Chen et~al.(2021)Chen, Burke, Hui, and Leykin]{chen2021understanding}
M.~Chen, R.~R. Burke, S.~K. Hui, and A.~Leykin.
\newblock Understanding lateral and vertical biases in consumer attention: an
  in-store ambulatory eye-tracking study.
\newblock \emph{Journal of Marketing Research}, 58\penalty0 (6):\penalty0
  1120--1141, 2021.

\bibitem[Chu(1989)]{chu1989paired}
C.~Chu.
\newblock A paired combinatorial logit model for travel demand analysis.
\newblock In \emph{Proceedings of the $5^{th}$ World Conference on
  Transportation Research, 1989}, volume~4, pages 295--309, 1989.

\bibitem[Corstjens and Doyle(1981)]{corstjens1981model}
M.~Corstjens and P.~Doyle.
\newblock A model for optimizing retail space allocations.
\newblock \emph{Management Science}, 27\penalty0 (7), 1981.

\bibitem[Corstjens and Doyle(1983)]{corstjens1983dynamic}
M.~Corstjens and P.~Doyle.
\newblock A dynamic model for strategically allocating retail space.
\newblock \emph{Journal of the Operational Research Society}, 34\penalty0
  (10):\penalty0 943--951, 1983.

\bibitem[Cox(1970)]{cox1970effect}
K.~K. Cox.
\newblock The effect of shelf space upon sales of branded products.
\newblock \emph{Journal of Marketing Research}, 7\penalty0 (1):\penalty0
  55--58, 1970.

\bibitem[Cressie(2015)]{cressie2015}
N.~Cressie.
\newblock \emph{Statistics for spatial data}.
\newblock John Wiley \& Sons, 2015.

\bibitem[Daly and S.(1976)]{daly1976zachary}
A.~Daly and Z.~S.
\newblock Improved multiple choice models.
\newblock In \emph{Proceedings of the Fourth PTRC Summer Annual Meeting.
  University of Warwick, England}, 1976.

\bibitem[Dotson et~al.(2018)Dotson, Howell, Brazell, Otter, Lenk, MacEachern,
  and Allenby]{dotson2018}
J.~P. Dotson, J.~R. Howell, J.~D. Brazell, T.~Otter, P.~J. Lenk, S.~MacEachern,
  and G.~M. Allenby.
\newblock A probit model with structured covariance for similarity effects and
  source of volume calculations.
\newblock \emph{Journal of Marketing Research}, 55 (February):\penalty0 35--47,
  2018.

\bibitem[Dreze et~al.(1994)Dreze, Hoch, and Purk]{dreze1994shelf}
X.~Dreze, S.~J. Hoch, and M.~E. Purk.
\newblock Shelf management and space elasticity.
\newblock \emph{Journal of Retailing}, 70\penalty0 (4):\penalty0 301--326,
  1994.

\bibitem[Hwang et~al.(2005)Hwang, Choi, and Lee]{hwang2005model}
H.~Hwang, B.~Choi, and M.-J. Lee.
\newblock A model for shelf space allocation and inventory control considering
  location and inventory level effects on demand.
\newblock \emph{International Journal of Production Economics}, 97\penalty0
  (2):\penalty0 185--195, 2005.

\bibitem[Koppelman and Wen(2000)]{koppelman2000paired}
F.~S. Koppelman and C.-H. Wen.
\newblock The paired combinatorial logit model: properties, estimation and
  application.
\newblock \emph{Transportation Research Part B: Methodological}, 34\penalty0
  (2), 2000.

\bibitem[Luce and Suppes(1965)]{luce1965utility}
R.~Luce and P.~Suppes.
\newblock Utility, preference and subjective probability.
\newblock \emph{Handbook of Mathematical Psychology}, 3:\penalty0 249--410,
  1965.

\bibitem[McFadden(1973)]{1973conditional}
D.~McFadden.
\newblock Conditional logit analysis of qualitative choice behavior.
\newblock \emph{Institute of Urban and Regional Development}, 1973.

\bibitem[McFadden(1978)]{1978modeling}
D.~McFadden.
\newblock Modeling the choice of residential location.
\newblock \emph{Transportation Research Record}, 1978.

\bibitem[McFadden(1981)]{mcfadden1981econometric}
D.~McFadden.
\newblock Econometric models of probabilistic choice.
\newblock \emph{Structural analysis of discrete data with econometric
  applications}, 198272, 1981.

\bibitem[McGranaghan et~al.(2019)McGranaghan, Liaukonyte, Fisher, and
  Wilbur]{mcgranaghan2019lead}
M.~McGranaghan, J.~Liaukonyte, G.~Fisher, and K.~C. Wilbur.
\newblock Lead offer spillovers.
\newblock \emph{Marketing Science}, 38\penalty0 (4):\penalty0 643--668, 2019.

\bibitem[Pauli and Hoecker(1952)]{pauli1952better}
H.~Pauli and R.~W. Hoecker.
\newblock \emph{Better Utilization of Selling Space in Food Stores: Part I,
  Relation of Size of Shelf Display to Sales of Canned Fruits and Vegetables},
  volume~30.
\newblock US Department of Agriculture, Production and Marketing
  Administration, 1952.

\bibitem[Peltzman(2000)]{peltzman2000prices}
S.~Peltzman.
\newblock Prices rise faster than they fall.
\newblock \emph{Journal of Political Economy}, 108\penalty0 (3), 2000.

\bibitem[Rabbani et~al.(2018)Rabbani, Salmanzadeh-Meydani, Farshbaf-Geranmayeh,
  and Fadakar-Gabalou]{rabbani2018profit}
M.~Rabbani, N.~Salmanzadeh-Meydani, A.~Farshbaf-Geranmayeh, and
  V.~Fadakar-Gabalou.
\newblock Profit maximizing through 3d shelf space allocation of 2d display
  orientation items with variable heights of the shelves.
\newblock \emph{Opsearch}, 55\penalty0 (2):\penalty0 337--360, 2018.

\bibitem[Rooderkerk et~al.(2011)Rooderkerk, Van~Heerde, and
  Bijmolt]{rooderkerk2011}
R.~P. Rooderkerk, H.~J. Van~Heerde, and T.~H. Bijmolt.
\newblock Incorporating context effects into a choice model.
\newblock \emph{Journal of Marketing Research}, 48 (August):\penalty0 767--780,
  2011.

\bibitem[Salomon and Ben-Akiva(1983)]{salomon1983use}
I.~Salomon and M.~Ben-Akiva.
\newblock The use of the life-style concept in travel demand models.
\newblock \emph{Environment and Planning A}, 15\penalty0 (5):\penalty0
  623--638, 1983.

\bibitem[Sayman et~al.(2002)Sayman, Hoch, and Raju]{sayman2002positioning}
S.~Sayman, S.~J. Hoch, and J.~S. Raju.
\newblock Positioning of store brands.
\newblock \emph{Marketing science}, 21\penalty0 (4):\penalty0 378--397, 2002.

\bibitem[Scekic et~al.(2018)Scekic, Atalay, Liu~Yang, and
  Ebbes]{scekic2018product}
A.~Scekic, S.~Atalay, C.~Liu~Yang, and P.~Ebbes.
\newblock Product search in retail environments: Influence of vertical product
  location on search performance.
\newblock \emph{ACR European Advances}, 2018.

\bibitem[Sener et~al.(2011)Sener, Pendyala, and Bhat]{sener2011accommodating}
I.~N. Sener, R.~M. Pendyala, and C.~R. Bhat.
\newblock Accommodating spatial correlation across choice alternatives in
  discrete choice models: an application to modeling residential location
  choice behavior.
\newblock \emph{Journal of Transport Geography}, 19\penalty0 (2):\penalty0
  294--303, 2011.

\bibitem[Small(1987)]{small1987discrete}
K.~A. Small.
\newblock A discrete choice model for ordered alternatives.
\newblock \emph{Econometrica: Journal of the Econometric Society}, pages
  409--424, 1987.

\bibitem[Smirnov and Huchzermeier(2019)]{smirnov2019shelf}
D.~Smirnov and A.~Huchzermeier.
\newblock Shelf-space management under stockout-based substitution and
  merchandising constraints.
\newblock \emph{Available at SSRN 3413256}, 2019.

\bibitem[Swait(2001)]{swait2001choice}
J.~Swait.
\newblock Choice set generation within the generalized extreme value family of
  discrete choice models.
\newblock \emph{Transportation Research Part B: Methodological}, 35\penalty0
  (7):\penalty0 643--666, 2001.

\bibitem[Vovsha(1997)]{vovsha1997application}
P.~Vovsha.
\newblock Application of cross-nested logit model to mode choice in tel aviv,
  israel, metropolitan area.
\newblock \emph{Transportation Research Record}, 1607\penalty0 (1):\penalty0
  6--15, 1997.

\bibitem[Wedel and Pieters(2008)]{wedel2008review}
M.~Wedel and R.~Pieters.
\newblock A review of eye-tracking research in marketing.
\newblock \emph{Review of Marketing Research}, 4\penalty0 (2008):\penalty0
  123--147, 2008.

\bibitem[Wen and Koppelman(2001)]{wen2001generalized}
C.-H. Wen and F.~S. Koppelman.
\newblock The generalized nested logit model.
\newblock \emph{Transportation Research Part B}, 35\penalty0 (7), 2001.

\bibitem[Williams(1977)]{williams1977formation}
H.~C. Williams.
\newblock On the formation of travel demand models and economic evaluation
  measures of user benefit.
\newblock \emph{Environment and Planning A}, 9\penalty0 (3):\penalty0 285--344,
  1977.

\end{thebibliography}
\vspace{-.5cm}
\appendix
\renewcommand{\theequation}{A\thechapter.\arabic{equation}}
\numberwithin{equation}{section}


\section{Appendix}

\subsection{Calculation of correlation using NCL} \label{ap:correlation}

To compute the induced correlation for a pair of products $i$ and $j$, from the NCL model we calculate a numerical covariance by numerical integration of their bi-variate density function. First, using the fact that the error term is distributed Gumble, we can construct the bi-variate marginal CDF for two products\footnote{To simplify the exposition, the only random component in the utility is the Gumble term, and the rest of the utility is equal.}, shown in equation \ref{bivariate_CDF}

\vspace{-1cm}
\begin{equation}
    \label{bivariate_CDF}
    H(\varepsilon_i,\varepsilon_j) = exp \biggl\{ -(1-\alpha_{i,ij})e^{-\varepsilon_i} -(1-\alpha_{j,ij})e^{-\varepsilon_j} - \left( (\alpha_{i,ij}e^{-\varepsilon_i})^{\frac{1}{\rho}}+(\alpha_{j,ij}e^{-\varepsilon_j})^{1/\rho}\right) \biggl\}
\end{equation}

The marginal probability density function associated with the CDF in equation \ref{bivariate_CDF} can be written as: $f(\varepsilon_i,\varepsilon_j)=H(\varepsilon_i,\varepsilon_j)\bigr[ B_{ij} C_{ij} + D_{ij} \bigr]$. Where:
\vspace{-.5cm}
\begin{align*}
    A_{ij} &=  (\alpha_{i,ij}e^{-\varepsilon_i})^{\frac{1}{\rho}}+(\alpha_{j,ij}e^{-\varepsilon_j})^{1/\rho} \\
    B_{ij} &= (1-\alpha_{i,ij})e^{-\varepsilon_i} + A_{ij}^{\rho-1} (\alpha_{i,ij}e^{-\varepsilon_i})^{\frac{1}{\rho}}\\
    C_{ij} &= (1-\alpha_{j,ij})e^{-\varepsilon_j} + A_{ij}^{\rho-1} (\alpha_{j,ij}e^{-\varepsilon_j})^{\frac{1}{\rho}}\\
    D_{ij} &= \left( \frac{1-\rho}{\rho} \right) A_{ij}^{\rho-2} (\alpha_{j,ij}e^{-\varepsilon_j})^{\frac{1}{\rho}}(\alpha_{i,ij}e^{-\varepsilon_i})^{\frac{1}{\rho}}\\
\end{align*}

\newpage

\begin{landscape}
\subsection{Summary Statistics}
\begin{table}[ht!]
\footnotesize
    \centering
\begin{tabular}{ccccccc}
    \multicolumn{4}{c}{Product} & \multicolumn{3}{c}{Average (Std Dev, Min, Max)} \\ \toprule
    $j$ & Name & Size & Market Share & Units Sold & Price ($\$$) & Profit ($\$$)\\ \hline
1 & CORONET SPARKLE-PRIN & 1 CT & 4.85\% & 103.55 (217.11, 1, 2172) & 0.79 (0.04, 0.69, 0.89) & 0.23 (0.06, 0.13, 0.40) \\ 
2 & HI DRI RECYCLED PP . & 1 CT & 3.67\% & 78.15 (105.76, 2, 1035) & 0.67 (0.04, 0.50, 0.69) & 0.19 (0.06, 0.02, 0.31) \\ 
3 & HI DRI PRINT PP.69 & 1 CT & 8.19\% & 174.58 (202.13, 11, 2400) & 0.67 (0.04, 0.50, 0.69) & 0.18 (0.06, 0.00, 0.29) \\ 
4 & BOUNTY WHIYE JUMBO & 1 CT & 5.31\% & 113.17 (194.07, 1, 3358) & 0.95 (0.08, 0.69, 1.06) & 0.15 (0.07, 0.00, 0.34) \\ 
5 & BOUNTY DESIGNER JUMB & 1 CT & 6.47\% & 138.11 (237.32, 2, 3533) & 0.95 (0.08, 0.69, 1.06) & 0.15 (0.07, 0.02, 0.33) \\ 
6 & BNTY BIG ROLL MCROWV & 1 CT & 2.45\% & 56.13 (30.83, 1, 240) & 1.47 (0.06, 1.31, 1.59) & 0.27 (0.09, 0.03, 0.44) \\ 
7 & BOUNTY BIG ROLL TWLS & 1 CT & 3.24\% & 73.49 (38.61, 1, 378) & 1.47 (0.06, 1.31, 1.59) & 0.27 (0.09, 0.03, 0.44) \\ 
8 & BOUNTY WHITE/DESIGNE & 3 CT & 2.11\% & 45.05 (19.78, 6, 126) & 2.82 (0.14, 2.59, 3.09) & 0.37 (0.15, 0.05, 0.74) \\ 
9 & DOM PAPER TOWELS & 1 CT & 11.14\% & 237.64 (341.18, 1, 3402) & 0.62 (0.04, 0.50, 0.69) & 0.14 (0.05, 0.02, 0.25) \\ 
10 & BOLT JUMBO DESIGNER & 1 CT & 0.60\% & 37.49 (28.01, 1, 379) & 0.92 (0.06, 0.86, 1.05) & 0.17 (0.05, 0.01, 0.27) \\ 
11 & BOLT DSNR TOWEL PP . & 1 CT & 1.16\% & 38.97 (33.02, 1, 503) & 0.88 (0.04, 0.69, 0.89) & 0.21 (0.04, 0.02, 0.22) \\ 
12 & BRAWNY BIG ROLL 50\% & 1 CT & 2.08\% & 44.30 (35.97, 1, 397) & 0.88 (0.05, 0.79, 0.96) & 0.24 (0.06, 0.14, 0.44) \\ 
13 & BRAWNY JUMBO DESIGNE & 1 CT & 1.99\% & 42.42 (32.06, 1, 346) & 0.88 (0.05, 0.79, 0.96) & 0.24 (0.07, 0.12, 0.46) \\ 
14 & BRAWNY 3 ROLL GIANT & 3 CT & 1.39\% & 29.77 (26.46, 1, 279) & 2.47 (0.18, 1.99, 2.73) & 0.50 (0.16, 0.11, 0.91) \\ 
15 & GALA DECORATIVE TOWE & 1 CT & 0.56\% & 25.16 (17.06, 1, 157) & 0.80 (0.05, 0.44, 0.86) & 0.22 (0.11, 0.01, 0.42) \\ 
16 & JOB SQUAD DECORATIVE & 1 CT & 3.23\% & 68.78 (31.53, 1, 293) & 0.88 (0.04, 0.80, 0.98) & 0.09 (0.04, 0.03, 0.25) \\ 
17 & SCOTTOWELS WHITE & 1 CT & 2.76\% & 59.05 (79.57, 1, 821) & 0.87 (0.05, 0.77, 0.96) & 0.17 (0.05, 0.06, 0.29) \\ 
18 & SCOTT TOWELS DECO & 1 CT & 3.71\% & 79.34 (106.93, 1, 1206) & 0.86 (0.05, 0.77, 0.96) & 0.15 (0.06, 0.01, 0.29) \\ 
19 & ~SCOTT TOWEL WHITE D & 3 CT & 2.44\% & 52.07 (45.39, 8, 527) & 2.52 (0.10, 2.32, 2.73) & 0.37 (0.13, 0.11, 0.80) \\ 
20 & SCOTT TOWELS ARTS & 1 CT & 4.02\% & 85.81 (111.58, 1, 1198) & 0.86 (0.05, 0.77, 0.96) & 0.15 (0.06, 0.02, 0.29) \\ 
21 & SCOTT MEGA PRINT TOW & 1 CT & 4.59\% & 97.88 (109.11, 1, 1629) & 1.29 (0.10, 0.99, 1.43) & 0.22 (0.10, 0.02, 0.48) \\ 
22 & VIVA ASSTD DEC TOWEL & 1 CT & 5.71\% & 121.75 (90.43, 7, 797) & 0.96 (0.05, 0.88, 1.09) & 0.17 (0.06, 0.00, 0.33) \\ 
23 & ~VIVA DECORATEDTOWEL & 1 CT & 5.42\% & 115.54 (88.21, 3, 750) & 0.96 (0.05, 0.89, 1.09) & 0.18 (0.06, 0.00, 0.33) \\ 
24 & GR FOREST TWL PP .69 & 1 CT & 4.55\% & 97.08 (76.62, 4, 847) & 0.69 (0.01, 0.65, 0.69) & 0.12 (0.02, 0.10, 0.22) \\ 
25 & MARDI GRA PP 2.29 TO & 3 CT & 1.44\% & 30.76 (22.78, 1, 213) & 2.26 (0.05, 2.09, 2.29) & 0.54 (0.09, 0.41, 0.81) \\ 
26 & MARDI GRA TWLS PP.79 & 1 CT & 6.94\% & 148.00 (199.69, 1, 1921) & 0.77 (0.04, 0.59, 0.79) & 0.19 (0.04, 0.02, 0.24)
\end{tabular}
    \caption{\footnotesize{Summary statistic for the data of the 26 products.}}
    \label{tab:ptw_desc}
\end{table}
\end{landscape} 

\end{document}